\documentclass[11pt,a4paper]{article}
\usepackage{geometry}  
\usepackage[british]{babel} 
\usepackage[T1]{fontenc} 
\usepackage[utf8]{inputenc}
\usepackage[dvips]{graphicx} 
\usepackage{url}
\usepackage{amsfonts}
\usepackage{hyperref}
\title{Combining local and global smoothing in multivariate density estimation}
\author{{\Large Adelchi Azzalini} \\  \large 
  \small University of Padua, Italy }
\date{\small\today}
\let\phi=\varphi
\newcommand{\Real}{\mathbb{R}}
\newcommand{\T}{^{\top}}
\newcommand{\pr}[2][]{
   \ensuremath{\mathbb{P}_{#1}\!\left\{\displaystyle{#2}\right\}}}
\newcommand{\inv}{^{-1}}
\newcommand{\diag}{\mbox{\rm diag}}
\def\biblioitem{\par\vskip\parskip\noindent\hangindent=2em \hangafter=1}

\begin{document}
\maketitle

\begin{abstract}\noindent
  Non-parametric estimation of a multivariate density estimation is tackled
  via a method which combines traditional local smoothing with a form of
  global smoothing but without imposing a rigid structure.  Simulation work
  delivers encouraging indications on the effectiveness of the method.  An
  application to density-based clustering illustrates a possible usage.
\end{abstract}


\section{Local and global smoothing} \label{s:intro}

Consider estimation of the probability density function $f(\cdot)$ of a
continuous random variable in cases when a parametric formulation for $f$ is
not considered appropriate.  
Given a random sample drawn form $f$, a variety of  non-parametric estimation 
methods are available.  Most of these methods share the common
feature of being `fully non-parametric', meaning that the set of competing
alternatives from which an estimate $\hat{f}$ must be selected is constituted
by the entire set of possible densities, except for some conditions of
mathematical regularity.

Limited work has been dedicated to methods which allow inclusion of some
qualitative requirement about $f$. One problem which has attracted a fair
amount of attention is estimation, in the univariate case, of a unimodal 
density or, more generally, of a density with a pre-assigned number of modes, 
like in Hall \& Huang (2002).
Other qualitative requirements on $f$ seem to have received less
consideration.

We shall be dealing with estimation of a  density $f(\cdot)$
on $\Real^d$, or possibly a subset of it, with $d>1$. 
For reasons which will become clear shortly, the case $d=2$ is
technically possible, but both uninteresting and nearly degenerate in our
framework; therefore $d\ge3$ is the situation really considered.  

It is well-known that, as $d$ increases, non-parametric methods, 
and in particular those 
for density estimation, degrade in performance, eventually running into 
the so-called problem of `curse of dimensionality' when $d$ is large. 
On the other hand, there is the widespread perception that, in many real
situations, the dependence structure of a multivariate distribution is
largely governed by the dependence among a smaller number of
components.  An explicit statement of this view has been expressed by Scott
(1992, p.\,195): ``Multivariate data in $\Real^d$ are almost never
$d$-dimensional.  That is, the underlying structure of data in $\Real^d$ is
almost always of dimension lower than $d$''.

The present contribution examines an estimation method motivated by these
considerations.  Broadly speaking, we impose a `light structure' on the
density $f(\cdot)$, moving away from a fully non-parametric construction, but
without imposing a detailed structure, such as parametric form.  A bit more
specifically, it is assumed that, at least in an approximate sense, the
dependence is regulated by a structure based on $m$-dimensional subsets of the
variables, with $m<d$.  The introduction of this constraint leads to a form of
global smoothing of the estimated distribution which can improve upon existing
methods, in appropriate situations, by reducing variability connected to
estimation of fine details of $f(\cdot)$ regulating high-order interactions
among variables. It is plausible that, even if these high-order interactions
are not exactly null, the reduction of variability of the estimate overcomes
the bias so introduced; an assumption of this sort is ubiquitous in any
modelling operation.
 
Clearly, the success of this scheme relies on the suitability of the imposed
structure in a given situation.  To exemplify by what here represents an
extreme case, application of the stated criterion when $d=2$ would entail to
introduce a joint distribution constituted only by marginals of dimension
$m=1$, that is, assuming independence of the two component variables.  In the
majority of situations, the more interesting ones in fact, this extreme
simplification would not be appropriate; this explains why earlier we have
effectively restricted ourselves to the case $d\ge3$.

Estimation of $f(\cdot)$ will still be carried out via a classical 
local smoothing method, such as the kernel estimator, but in a way 
which reflects the global smoothing imposed  by the assumed structure 
of $f(\cdot)$. Therefore the final outcome of the procedure
will reflect both the local and the global smoothing operations.

In the next two sections, we transfer this broad criterion into a
specific operational formulation. This is then followed by numerical
exploration to evaluate its practical working with simulated data and by its 
utilization within a density-based clustering process of some real data.


\section{Global smoothing via a log-linear model}  \label{s:loglin}

The criterion described only qualitatively so far can be translated 
into an operational procedure. Given the broad nature of the
above formulation, there is not a unique prescribed way 
to  define such a  procedure. The route to
be presented here is driven by simplicity and flexibility, since it can
be used in conjunction with  any local smoother which allows weighted
observations with only simple adjustments of an existing method.

Assume that a sample $z_1, \dots, z_n$ of  observations drawn from 
$f(\cdot)$ is available, where $z_i= (z_{i1}, \dots, z_{id})\T$ for
$i=1,\dots, n$. Denote by $Z=(Z_1, \dots, Z_d)\T$ the parent random
variable from which the $z_i$'s are drawn, all independently from
each other. 
We introduce subdivisions of the $d$ coordinates axes into disjoint 
$r_1,\dots,r_d$  intervals, creating a partition of the sample
space into $r=r_1\times\cdots\times r_d$ hyper-rectangles or cells.
Correspondingly, there are probabilities $\pi_1,\dots, \pi_r$ 
associated to the cells. 
The $j$th element of this partition, denoted $R_j$, can be associated
to a multidimensional subscript $j_1,j_2,\dots, j_d$, but this is not
of relevance at the moment. 

Denote  by $n_j$ the number of sample elements falling into $R_j$, 
for $j=1,\dots, r$, so that $\sum_j n_j=n$. The basic estimate of
$\pi_j=\pr{R_j}$ is $n_j/n$ and, correspondingly, for a given point 
$x\in R_j$, a crude estimate of $f(x)$ is 
\begin{equation}
    \frac{n_j}{n\;\mathrm{vol}(R_j)} 
    \label{e:crude-f^}
\end{equation}
where  $\mathrm{vol}(R_j)$ is the geometric volume of $R_j$.

A key weakness of this scheme is that it implies
$r-1$ distinct probabilities $\pi_j$ to be estimated, 
up to a constraint on their sum. If $d$ is not
small and the $r_1,\dots,r_d$ subdivisions are not coarse,
$r$ can be large. To reduce the number of
free parameters to be estimated, we introduce a log-linear model
for the cells probabilities, $\pi_j$'s, where interaction terms
involving more than $m$ component variables are set to zero.
For instance, if $d=3$ and we choose $m=2$, terms of the
log-linear representation of the $\pi_j$'s are retained 
up to pairwise  interactions while all three-factor
interaction terms are eliminated, reducing the number of 
underlying parameters by $(r_1-1)\,(r_2-1)\,(r_3-1)$ with 
respect to the saturated model.
See Section~9.2.2 of Agresti (2013) for a detailed discussion
of the pertaining log-linear model; there is only the
difference that  those expressions refer to the expected 
values of the frequencies instead of the probabilities, 
but this is irrelevant since the two sets of quantities 
are proportional to each other.

After the log-linear model has been fit to the observations, 
a set of expected frequencies is obtained, denoted 
$\hat{n}_j$, and corresponding estimated probabilities
$\hat{\pi}_j \; (j=1,\dots,s)$.
Replacing $n_j$ by $\hat{n}_j$ in (\ref{e:crude-f^}) provides
a revised estimate. 

The constraints enforced by the log-linear model refer to the
cell probabilities, $\pi_j$, and so to the expected frequencies
$\hat{n}_j$, but not to the density $f(x)$. 
However, provided the values $\mathrm{vol}(R_j)$ are not too disparate, 
at least in the region where most of distribution is located, 
the originally intended dependence structure will hold approximately. 
Furthermore, the local smoothing step to be presented shortly
introduces an additional perturbation in this sense.
Since the imposed dependence structure is motivated by 
practical considerations of improved estimation performance 
rather than exact model compliance, we are not concerned about
these approximations. 

\section{Non-parametric local smoothing}  \label{s:wkde}

In the second step of the procedure, the expected frequencies $\hat{n}_j$ 
obtained in the first step are used to assign weights to the observations, 
$z_i$, so that the estimate $\hat{f}$ produced by the subsequent local 
smoothing respects, approximately, the estimated cell probabilities 
$\hat{\pi}_j$.

To illustrate the procedure, we use the following simple form
of the  kernel density estimate.
Take the kernel function to be the
$d$-dimensional circular normal density with standardized components, 
denoted $\phi_d(\cdot)$, and choose a vector $h_1, \dots, h_d$ of
smoothing parameters; then the classical kernel estimate at point
$x\in\Real^d$  is
\begin{equation}
   \frac{1}{n\,\det(h)}\:\sum_{i=1}^n \phi_d\left( h\inv (x-z_i)\right)
   \label{e:kde-classical}
\end{equation}
where $h=\diag(h_1,\dots, h_d)$. This classical estimate is modified
by weighting an observation in the $j$-th cell with $w_j=\hat{n}_j/n_j$, 
so that the overall weight of the $n_j$ observations in the $j$th cell is 
$\hat{n}_j$ instead of $n_j$. The new estimate takes the form
\begin{equation}
   \hat{f}(x) = \frac{1}{n\,\det(h)} \:\sum_{i=1}^n 
                w_{j(i)}\: \phi_d\left( h\inv (x-z_i)\right)
   \label{e:kde-weighted}
\end{equation}
where $w_{j(i)}$ is the weight of the cell to which $z_i$ belongs.
The type of perturbation of estimator (\ref{e:kde-classical}) is denoted 
`tilting' by Doosti \& Hall (2016). 

Clearly several variants forms can be considered, such as replacing the normal
kernel in (\ref{e:kde-weighted}) by some other multivariate kernel or using
smoothing parameters which vary with the observations.  Not only these
variants are immediately accommodated, but we are not restricted to
kernel-based methods. For instance, if we use instead a projection method
based on an orthogonal series decomposition of $f$, the weight $w_{j(i)}$ is
assigned to observation $z_i$ when the coefficients of the projection are
estimated as sample averages of suitable data-dependent functions.

A complication arises with empty cells, where $n_j=0$, since the corresponding
weights $w_j$ are not well defined. While for cells with $n_j>0$ the method
works by suitably increasing or decreasing the weight of the observations
belonging to that cell, no such adjustment is possible if the cell is empty.
 
The simplest approach to the problem is to just use (\ref{e:kde-weighted}), 
only replacing $n$ in the denominator by $n_w=\sum_i w_{j(i)}$; we are then
effectively ignoring the `empty cells' problem.  This variant form is
denoted `Plain' (or P) later on. When $\hat{n}_j$ corresponding to $n_j=0$ 
is of non-negligible magnitude, possibly so for several cells, the
above solution may appear too crude, hence supporting the alternative approach
of introducing some fictitious data to fill the empty cells. Note that 
this implies that the overall number of data points exceeds $n$, although 
the sum of their weights remains $n$.  
However, while appealing in principle, it is hard to say how to pursue 
this route in a theoretically-motivated procedure.  An heuristic method 
has been constructed, described in Appendix~A.
This variant form is denoted `Fill' (or F) later on.

\section{Practical and computational aspects}

The practical implementation of the method requires to specify,
in the first place, the subdivisions of the
axes which identify the $r=r_1\times\cdots\times r_d$ cells. For each axis, 
we have started by applying the `normal reference rule' proposed by 
Scott (1992, p.\,82) for choosing the histogram bins,  
assuming joint normality of the multivariate  distribution, that is,
the $k$th bin-width is initially taken equal to 
\[
   b_k \approx 3.5\,\sigma_k \: n^{-1/(2+d)}
\] 
where $\sigma_k$ is the standard deviation of the $k$-th component
variable, $Z_k$; in practice, $\sigma_k$ must be replaced by its sample
value. Division of the range of observations $(z_{1k},\dots, z_{nk})$ by
$b_k$, rounded to the nearest integer, lends the number of bins for the $k$-th
component, $r_k$.  The sample range of the $k$-th variable has then been
subdivided into $r_k$ intervals, constructed as follows: first the quantiles
of level $(0, 1/r_k, \dots, r_{k-1}/r_k,1)$ of the $\mathrm{Beta}(3/4, 3/4)$
distribution have been computed and then the sample quantiles with level equal
to these Beta quantiles have been used as the end-points of the intervals on
the $k$-th axis.  The underlying idea is to have the central intervals 
 shorter than those near the margins of the sample range.  
This process is repeated for $k=1,\dots,d$.

This procedure for choosing the end-points of the intervals appears somewhat
arbitrary if examined from a formal viewpoint.  The scheme must rather be
regarded as a way of mimicking the non-automatic process followed when
intervals are chosen by subjective judgement.

To compensate the possible effect of the choice of the number of 
subdivisions $r_1,\dots, r_d$, a variant form of the procedure involved 
using three choices  of the subdivisions: one as described
above, one decreasing each $r_k$ by 1 and the third one 
increasing each $r_k$ by 1. For each of these three grids, the log-linear
fitting and computation of  (\ref{e:kde-weighted}) were applied, followed
by averaging of the three estimates. This variant is denoted `Average' (or A)
later on.

Once the grid of the $\Real^d$ space has been fixed, the sample frequencies
$n_j$ of the cells are identified. 
For a given value of $m$, we must fit a  log-linear model as described 
in Section~\ref{s:loglin}; we temporarily leave aside the choice of $m$,
to which we return later. In our problem the interest is
only in the fitted frequencies, $\hat{n}_j$, not in the log-linear parameters.
In this case the recommendation of Agresti (2013, \S\,9.7.3) is to adopt the
iterative proportional fitting algorithm, since it ``converges to the ML
[maximum likelihood] fit even when the likelihood is poorly behaved''.  A
Fortran implementation of this algorithm has been provided by Haberman (1972),
subsequently ported to the \texttt{R} computing environment with name
\texttt{loglin}.

The final step is application of the the weighted estimator
(\ref{e:kde-weighted}).  In most of our numerical work, the diagonal smoothing
matrix $h$ has been chosen by the multivariate version of the plug-in method
of Wand and Jones (1994) available in the \texttt{R} package \texttt{ks}
(Duong, 2015).

\begin{figure}
\centerline{\includegraphics[width=0.7\hsize]{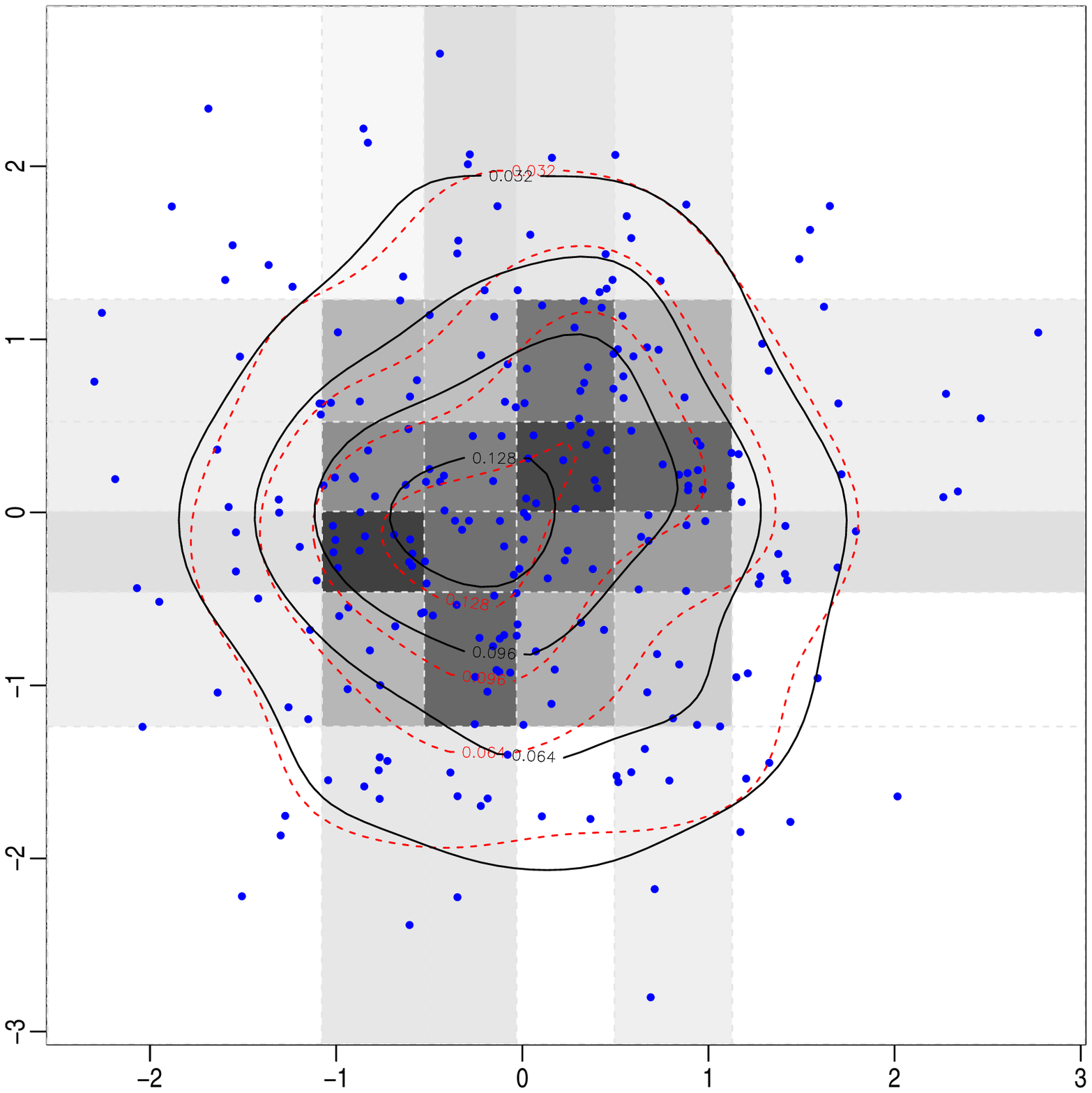}}
\caption{Illustrative example of the estimate in the simple case with $d=2$
  and $m=1$}
\label{fig:lls-2D-example}
\end{figure}

An illustration of the working of the procedure is
provided in Figure~\ref{fig:lls-2D-example}, which refers to the simplest
possible case, that is, with $d=2$ and $m=1$. As already explained, this
situation is not of practical relevance, but it is appropriate for simple
illustration.  Specifically, the $n=250$ plotted points constitute a sample
drawn from a circular bivariate normal distribution with standardized
marginals.  A rectangular area slightly wider than the range of the observed
points has been selected and, using the above-described rule, a $6\times6$
grid has been identified; for all cells of this grid, $n_j>0$ was observed.
The $36$ rectangles have been shaded using a $25$-level grey scale which
discretizes the values of the crude estimate (\ref{e:crude-f^}).  The dashed red
lines represent the contour level curves of the classical kernel estimate
(\ref{e:kde-classical}) while the continuous black curves refer to the weighted
estimate (\ref{e:kde-weighted}); this estimate appears somewhat smoother than the
unweighted one, with more limited departures from convexity, especially so in
the central region.

\section{Simulation work} \label{s:simulation}

The performance of the proposed method has been examined in a 
number of cases, using simulated data from a range of distributions: normal,
skew-normal, Student's $t$ and its skew version, and two-component mixtures 
of these distributions. The general expression of the distribution in use is
\begin{equation}
 f(x) =  f(x) = \pi f_1(x) + (1-\pi) f_2(x), \qquad x\in\Real^d,
 \label{e:f(x)-mixture}
\end{equation}
where $f_1$ and $f_2$ are of skew-normal (SN) or skew-$t$ (ST) type, which
include the classical normal and $t$ distributions as special cases;
$\pi\in(0,1]$ is the mixing proportion. The distributions $f_1$ and $f_2$ are
specified by the following parameters: a $d$-vector location $\xi$, a $d\times
d$ symmetric positive-definite scale matrix $\Omega$, a $d$-vector slant
$\alpha$ and a positive real number $\nu$.  The component $\nu$ exists only
for the ST distribution; when $\nu=\infty$, or equivalently when it is not
present, the distribution is of SN type.  A detailed treatment of the
multivariate SN and ST distributions is provided by Azzalini \& Capitanio
(2014).  When $\pi=1$, there is effectively no mixture mechanism and only the
$f_1$ parameters are required. The parameters considered have been selected
among the following options.

\begin{itemize}
\item If $\pi=1$, the location parameter is always $\xi=(0,\dots,0)\T$. 
  If $0<\pi<1$, the location of $f_1$ is $2\times1_d$ and
  the one of $f_2$ is $-2\times 1_d$, where $1_d$ denotes the $d$-vector of all
  1's.
\item The scale matrix $\Omega$ has been chosen among the following options:
 \begin{itemize}
  \item the identity matrix $I_d$;
  \item a Toeplitz-type matrix with $(i,j)$-th entry $\rho^{|i-j|}$ 
    where $\rho=3/4$,  or equivalently with AR(1) correlation structure;
  \item an ARMA(2,1) correlation structure;
  \item a matrix with elements specified individually, in some instances with
    $d=3$.   
  \end{itemize}  
\item The components of the slant parameter $\alpha$ have been specified 
  individually. When  $\alpha=0$ and $\pi=1$, the distribution is a regular
  (symmetric) Gaussian or Student's $t$ distribution.
\item The degrees of freedom $\nu$ was given a value among the following: 
  $\infty$ (corresponding to the SN distribution), 5 or 2.
\end{itemize}   

Distributions with dimension from $d=3$ to $d=5$ have been considered. 
The value $m$ employed in the log-linear model was $m=2$ and $m=3$,
with the constraint $m<d$.
In most cases, the sample size was $n=500$; 
a few experiments used either $n=250$ or $n=1000$. 
For each  combination of parameter values and sample size, 
$2500$ replicates have been considered and the following estimation 
methods have been tested: 
\begin{enumerate} 
  \setlength{\leftmargin}{1ex}
   \setlength{\parskip}{0ex} \setlength{\itemsep}{0ex} 
   \setlength{\topsep}{0ex} \setlength{\partopsep}{0ex}
\item the classical kernel method in (\ref{e:kde-classical}), denoted `kde';
\item the weighted kernel method in (\ref{e:kde-weighted}), in the Plain 
 variant form described in Section~\ref{s:wkde};
\item the Average variant which averages three estimates computed from
 three grid subdivisions of the sample space;
\item  the Fill variant which  places constructed points in empty cells,
  as described at the end of Section~\ref{s:wkde} and more in detail in
  an Appendix.
\end{enumerate}
The vector of smoothing parameters for the kernel method and its variants 
was obtained by the function \texttt{Hpi.diag} of the \texttt{R} package 
\texttt{ks} described earlier.
 
The possible number of factor combinations so obtainable is enormous even if
one selects only a few possible choices for each of the above-described
parameters.  Moreover the computation burden with certain parameters,
especially for the Average variant, was appreciable.  This prevents any
attempt of running a full factorial experiment.  Only a selection of factor
combinations has been considered, driven by subjective judgement on the
outcome of earlier experiments, paying more attentions to situations which
appeared more interesting in some sense. For instance, for a certain
combination of parameters, the value of $n$ could have been decreased to $250$
or increased to $1000$ to examine the effect of sample size alone, when this
appeared to be an interesting case.

To evaluate the performance of the proposed estimate (\ref{e:kde-weighted}), in
their variant forms, with respect to the classical estimate
(\ref{e:kde-classical}), the estimation error at $x\in\Real^d$ has been
expressed by
\begin{equation}
  e(x) = \frac{|f(x)- \hat{f}(x)|}{f(x)^p}   \label{e:e(x)}
\end{equation}
where $p=1/2$ has been used in the outcomes presented below. Initial numerical
work had also considered $p=0$ and $p=1$, but the general qualitative
indication which emerged was not very different and $p=1/2$ may represent a
reasonable compromise between absolute error and relative error.  Two sets of
points have been considered for evaluating (\ref{e:e(x)}): (i)~a non-random grid
of points spanning the area of non-negligible density of $f(x)$ and (ii)~the
sample values.  The second option is relevant in certain applications like the
one of Section~\ref{s:cluster}. A detailed description of the non-random grid
of points is provided in Appendix B. For the (ii) case, only the real
observations have been considered in the Fill variant, ignoring the fictitious
observations which it involves.

\begin{figure}\vskip-4ex
\begin{center}
\includegraphics[height=0.3\textheight,width=0.7\hsize]{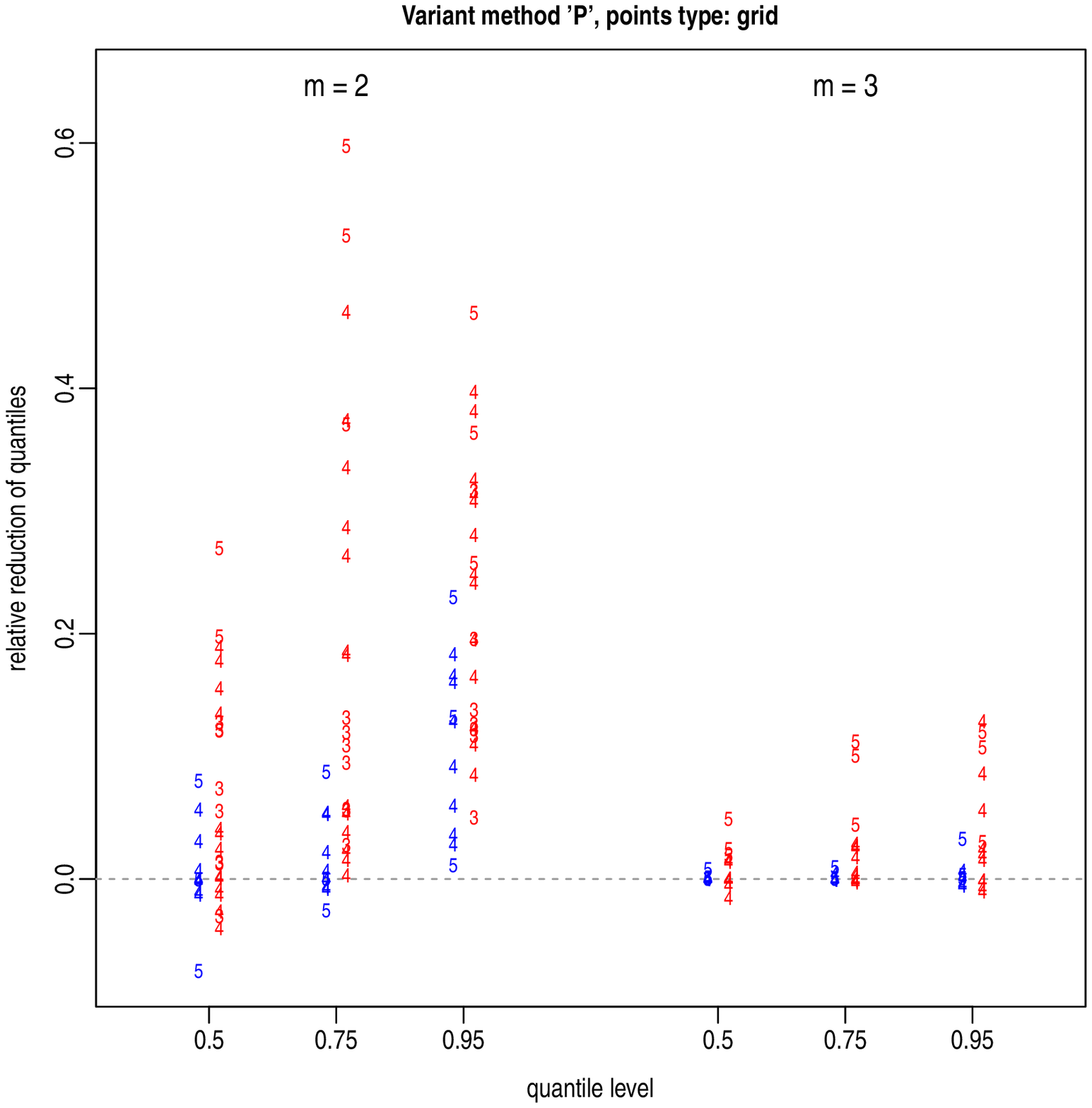}
\par\vspace{2ex} 
\includegraphics[height=0.3\textheight,width=0.7\hsize]{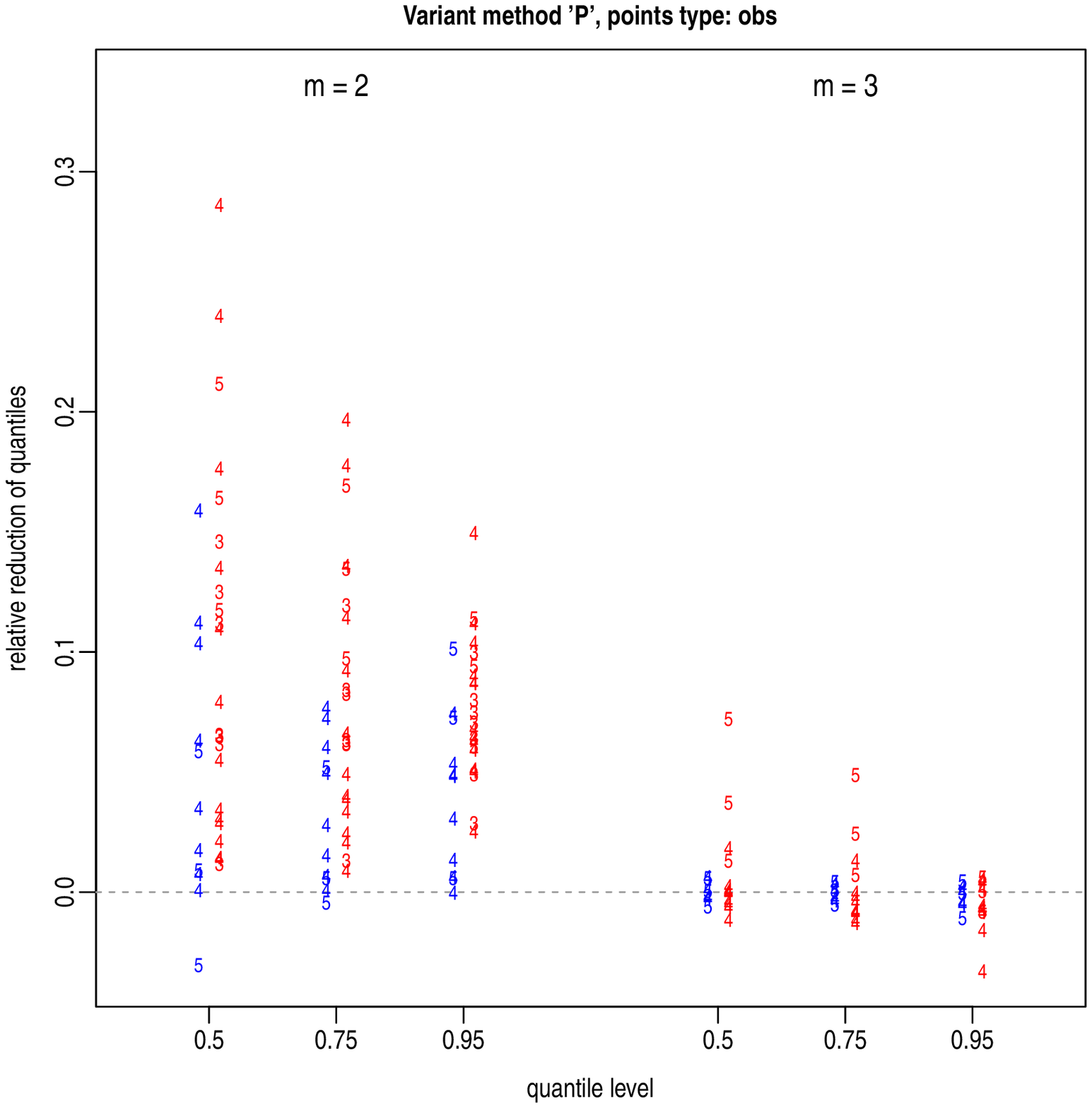}
\par\vspace{2ex}
\includegraphics[height=0.3\textheight,width=0.7\hsize]{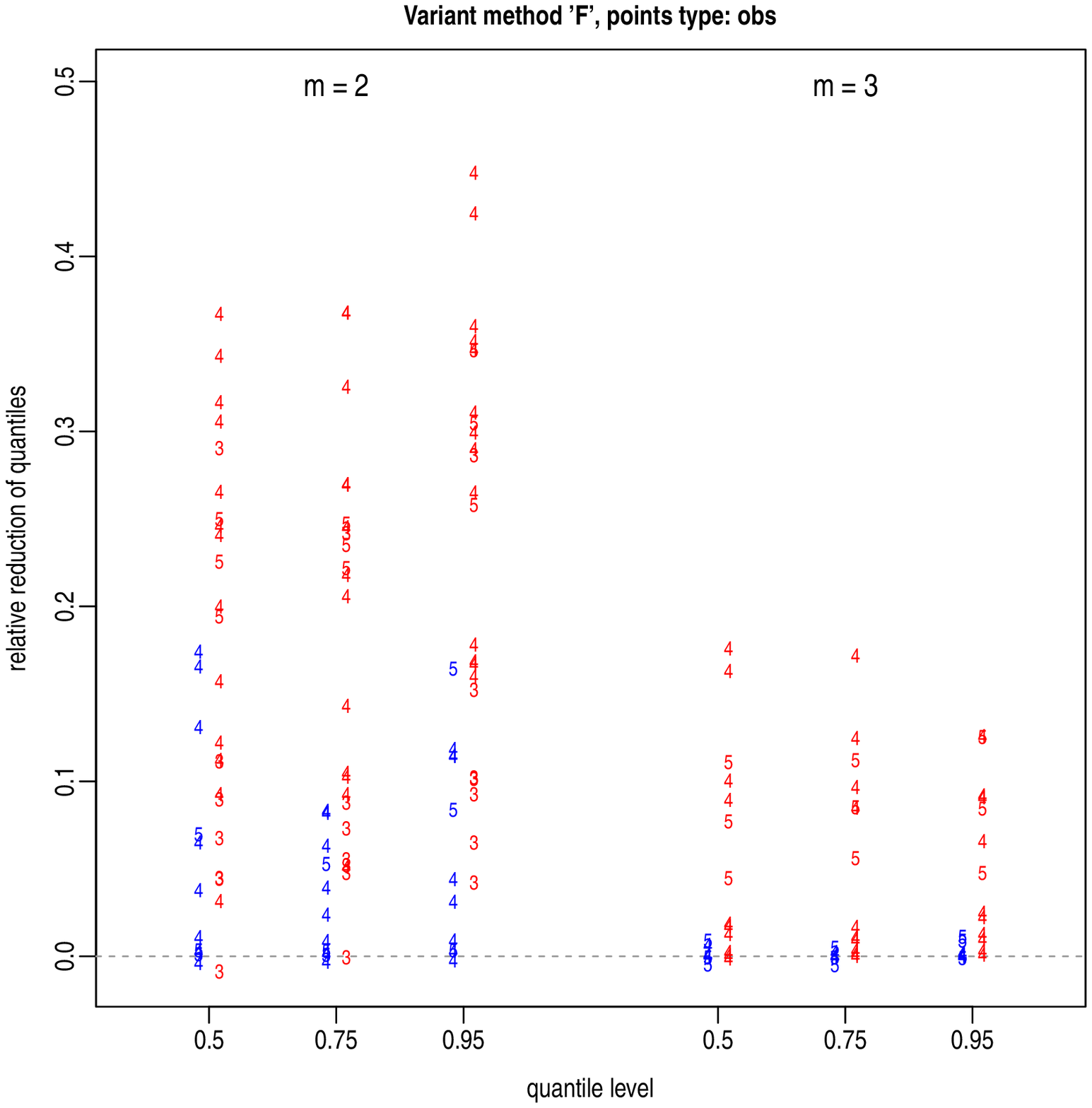}
\end{center}
\caption{Relative reduction of estimation error quantiles of the proposed
  estimate versus the classical kernel estimate; see text for a detailed
  description.}
\label{fig:rel-improv}  
\end{figure}

Direct consideration of  (\ref{e:e(x)}) for all the evaluation points, 
in either of the two considered sets, is not feasible. The quantiles of such 
sets of estimation errors have examined instead, at probability levels
$(0.25, 0.50, 0.75, 0.90, 0.95, 0.99)$. Even with this reduction, the amount
of tabular material so produced is considerable; the full set of
such tables is provided  in Appendix B.
A more compact summary exhibit of the overall outcome is provided by 
Figure~\ref{fig:rel-improv}. The values on the vertical axis represent 
\begin{equation}
    R(p) = \frac{Q_0(p) - Q(p)}{Q_0(p)}  \label{e:improve.Q(p)}
\end{equation}
where $Q(p)$ is the $p$-level quantile of the relative error (\ref{e:e(x)}),
evaluated over a given set of points,
for a the proposed method (in one of its variants) and $Q_0(p)$ is the 
similar quantity for the standard kernel density estimation.  
Therefore, $R(p)$ represents a measure of  reduction 
of the estimation error with respect to the classical estimate, 
or a  measure of its increase in case this quantity is negative.
Figure~\ref{fig:rel-improv} reports only the more noteworthy aspects 
of the full outcome, as described next. 

Only variants Plain and Fill have been considered in
Figure~\ref{fig:rel-improv} since the Average form was essentially equivalent
to the Plain one, with extra computing effort; it is however reassuring to
know that the specific choice of the grid size is not critical.  For each of
the three panels of the figure, the left portion refers to the choice $m=2$
for the log-linear model, the right portion to $m=3$.  Three values of $p$
entering (\ref{e:improve.Q(p)}) have been reported, namely $p=0.50, 0.75, 0.95$,
from the full six values in the complete outcome.  For each pair of $p$ and
$m$, there are two vertical stripes of numbers; the left blue stripe refers to
distributions which are mixtures, while the right red stripe refers to
single-component distributions, that is, those having $\pi=1$ in
(\ref{e:f(x)-mixture}).  In all cases, the digit plotted at ordinate
(\ref{e:improve.Q(p)}) denotes $d$.  Of the three panels, two refer to the Plain
variant of the method, with evaluation is performed either at a fixed grid of
points or at the sample points. The third panel refers to the Fill variant,
but only with evaluation at the sample points; evaluation at the grid points
was markedly unsatisfactory.
  
The first message emerging from inspection of Figure~\ref{fig:rel-improv} is
that an improvement of the weighted kernel estimate over the classical 
one occurs in the majority of cases, often with an appreciable magnitude; 
the negative values are limited in number and in magnitude. 
This consideration is substantially reinforced if we confine
attention to $m=2$, irrespectively of $d$; this explain why $m>3$ has not
been considered in the simulations. Another indication is that
the method, in all variants, performs better with a single component
distribution than with a non-generate mixture.

Operationally, the following recommendations for use of the method can be 
extracted: (i) set $m=2$ in all cases; (ii) the Plain variant is preferable
when the whole density surface must be estimated, while the Fill variant
is preferable for evaluation at the observed data points;
(iii) expect more improvement in case of a unimodal distribution than
a multimodal one. These recommendations refer to the kernel estimate 
and smoothing matrix described above, and they may not necessarily hold 
for other forms of non-parametric estimation.

\section{Application to density-based clustering}  \label{s:cluster}

The proposed density estimate has been used in conjunction with the
clustering method presented by Azzalini and Torelli (2007), implemented
in the \texttt{R} package \texttt{pdfCluster} (Azzalini and Menardi, 2014).
Since this clustering technique is firmly based on estimation of the 
density of the underlying $d$-dimensional random variable, it represents 
an ideal framework for application of the present proposal.

The real-data application presented  by Azzalini and Torelli (2007, 
Section 4.3)  concerned eight chemical components of $n=572$ specimens 
of olive-oil originating from various regions of Italy.
We re-examine their clustering exercise whose aim was the reconstruction 
the production area of the specimens from the values of their
chemical components.
The data themselves are available in the \texttt{pdfCluster} package.
A more detailed description of the data and of their preliminary 
transformations, which we also apply here, is provided by Azzalini and 
Torelli (2007).  We only specify the undocumented detail that, 
in the  additive log-ratio transform  applied to the compositional data 
$p_j$ ($j=1,\dots,8$), 
 namely $y_j=\log(p_j/p_k)$  for $j\not=k$, the  choice $k=4$ was made
because the values of $p_4$  are well separated from 0; this is also the
choice of the original article.  The first $d=5$ principal components of the 
$y_j$'s constitute the variables used for the actual clustering step.

The \texttt{pdfCluster} package was applied to the five principal components
just described  both in its current public version  
(1.0-2, as available at the time of writing) and a modified version
which replaces the classical kernel estimate (\ref{e:kde-classical})
and the weighted kernel estimate (\ref{e:kde-weighted}) with $m=2$;
all other ingredients have been kept at the default
specification of the package.  Table~\ref{tab:cluster} displays the 
cross-classification table of the true geographical areas and the groups
formed by clustering for the classical estimate 
(\ref{e:kde-classical})  in the first three columns and the new proposed
estimate in the last three columns refer. The latter estimate has 
actually been computed using both the Fill and the Plain variant,
but the outcome was the same. 
The ARI values underneath each sub-table denote the
`adjusted Rand index' which constitutes a measure of agreement between the
true and the reconstructed classification (Huber and Arabie, 1985). There is a
clear improvement in using the new estimate, from consideration both of direct
inspection of the table and by the ARI values.  
The left portion of Table~\ref{tab:cluster} is slightly different from the 
table originally obtained by Azzalini and Torelli (2007), but the essential 
traits are the same and the ARI value was even smaller there, namely 0.792.

\begin{table}[ht]
\centering
\caption{Clustering of olive-oil data: true versus reconstructed groups
 using the current package  \texttt{pdfCluster} and its modification with
 density estimation replaced by the new method.}
 \label{tab:cluster}
\vspace{1ex} 
\begin{tabular}{l*7r} 
  \hline 
  & \multicolumn{3}{c}{classical estimate} && 
     \multicolumn{3}{c}{proposed estimate}\\
   \hspace{8em}  & 1 & 2 & 3 & \hspace{2em}  & 1 & 2 & 3\\ 
  \hline
   South         & 321 &  0 &  2  &  & 323 &   0 &   0 \\ 
   Sardinia      &  0  & 98 &  0  &  &   0 &  98 &   0 \\ 
   Centre-North  &  0  & 45 & 106 &  &   0 &  22 & 129 \\ 
   \hline
   ARI & \multicolumn{3}{c}{0.873} &&  \multicolumn{3}{c}{0.937}
\end{tabular}
\end{table}

The values in  Table~\ref{tab:cluster} have been obtained using the
default smoothing parameter $h$ of  \texttt{pdfCluster}, which is
the asymptotically optimal bandwidth under normality,  multiplied 
by a shrinkage factor.
For completeness, we considered also the choice of $h$ produced by 
\texttt{Hpi.diag}, already used in the simulation work. 
In this case the shrinkage factor usually introduced by  
\texttt{pdfCluster} has  been  to the neutral value of 1,
since that shrinkage loses meaning with another choice of $h$.
The ARI value of the new groupings decreases
slightly to 0.910 for the proposed estimate, while where was
a much worse degrade for the classical estimate, which lead to four 
groups instead of three, with an ARI of 0.817.

At first sight, it may look surprising that the use of estimate
(\ref{e:kde-weighted}) produces such a noticeable improvement over
(\ref{e:kde-classical}), considering that Figure~\ref{fig:rel-improv} indicates
a limited improvement in connection with multimodal densities, which is the
typical situation in a clustering context. One must however bear in mind that
the procedure underlying \texttt{pdfCluster} involves two main stages: in the
first stage, the density of the overall population is estimated, to
locate the cluster cores associated to the subpopulations, while, in the
second stage, the distribution of each identified cluster core is estimated
separately.  
The densities of these sub-populations are naturally of unimodal type, 
where Figure~\ref{fig:rel-improv} indicates a better performance.  
It is then reasonable to link the successful effect of the new 
estimate mainly to its role in the second stage of the procedure.
 
\section{Final remarks}
The numerical outcome, both from the simulation work and from the clustering 
application, provides quite clear evidence in support of the proposed method.
However, there is still much room for improvement. For instance, a better
motivated method for filling empty cells would be welcome.
Even more importantly, some mathematically-argumented understanding  of 
why the method works is lacking.  Moreover, the global smoothing technique
of Section~\ref{s:loglin} represents one possible route to implement the
qualitative criterion stated in Section~\ref{s:intro}, but other routes 
may be considered.

An implementation of the proposed method will be made publicly available
in the \texttt{R} package \texttt{pdfCluster}.
\paragraph{Acknowledgements} The development of this work has much benefited 
from stimulating discussions with Giuliana Regoli.
\clearpage

\section*{References}
\biblioitem
Agresti, A. (2013). \emph{Categorical Data Analysis}, 3rd edition.
  Wiley, New York.

\biblioitem
  Azzalini, A. with the collaboration of Capitanio, A. (2014).
  \emph{The Skew-Normal and Related Families}. 
  Cambridge University Press,  IMS monographs. 
  
\biblioitem  
  Azzalini, A. and Menardi, G. (2014). Clustering via nonparametric density estimation:
  The \texttt{R} package \texttt{pdfCluster}. 
  J. Stat. Software, \textbf{57}(11), 1--26. \url{http://www.jstatsoft.org/v57/i11/}

\biblioitem
  Azzalini, A and Torelli, N. (2007). 
  Clustering via nonparametric density estimation.
  \emph{Stat. Comput.}, \textbf{17}, 71--80.

\biblioitem
  Doosti, H. and Hall, P. (2016)
  Making a non-parametric density estimator more attractive, and more accurate,
  by data perturbation.
 \emph{J.\ R.\ Stat.\ Soc, series B}, \textbf{78}, 445--462.


\biblioitem
  Haberman, S. J. (1972).
  {Algorithm AS 51}: Log-linear fit for contingency tables. 
  \emph{Appl. Stat.}, \textbf{21}, 218--225.

\biblioitem
  Hall, P. and Huang, L. S. (2002).
  Unimodal density estimation using kernel methods.
  \emph{Stat. Sinica}, \textbf{12}, 965--990.

\biblioitem
  Scott D. (1992).
  \emph{Multivariate Density Estimation}. John Wiley \& Sons, New York. 
%

\biblioitem
  Duong, T. (2015). \texttt{ks}: Kernel Smoothing. \texttt{R} package version 1.10.0.
  \url{https://CRAN.R-project.org/package=ks}

\biblioitem
  Wand, M.\,P. and Jones, M.\,C. (1994). 
  Multivariate plugin bandwidth selection. \emph{Comp. Stat.} \textbf{9}, 97--116.

\clearpage
\appendix

\section*{Appendix}
\subsection*{A. Filling empty cells}
As explained in Section~\ref{s:wkde}, empty cells having $n_j=0$ are
problematic.  One approach is to fill them with some fictitious data before
applying the weighted kernel estimate (\ref{e:kde-weighted}). Unfortunately, the
construction of such data by some theoretically-supported procedure appears to
be a challenging problem. We describe instead a fairly simple heuristic
procedure.
 
Consider a given cell $R_j$ with  $n_j=0$ but $\hat{n}_j>0$ and denote 
by $\bar{n}_j= \lceil n_j\rceil$ 
the smallest integer value larger than or equal to $\hat{n}_j$. 
The aim is to choose  $\bar{n}_j$ fictitious
points in $R_j$; recall that the points will be suitably weighted so that the
overall weights of the cell will be $\hat{n}_j$. 

An instinctive idea is to consider a component-wise average of the coordinates
of some neighbouring observations falling in adjacent cells, but this may
easily produce points outside the cell $R_j$.  To avoid this problem, we
consider instead an average of the coordinates of the corners of $R_j$, 
giving more weight to the corners closer to nearby observations.  
The specific procedure is as follows:
\begin{itemize}
\item the set of Euclidean distances of each observations from the centre of
  $R_j$ are computed and sorted in increasing order;
\item construct a fictitious observation from the following two steps:  
\begin{enumerate}
\item compute the Euclidean distances between the first element of the
  available observations (in the sorted list just constructed) and the $2^d$
  corner points of $R_j$, and assign to each corner point a weight inversely
  proportional to the square root of its distance;
\item the above step is repeat $n_a$ times (in our work $n_a=3$ has been
  used), each time discarding the already employed observation from the sorted
  list, adding up the weights of the corner points; finally, retain the
  weighted average of the corner points as a new constructed observation;
  \end{enumerate}   
\item step 1 and 2 are repeated for each of the $\bar{n}_j$ points to be
  constructed.
\end{itemize}

\subsection*{B.~Output of the simulation study}

The definition of the distributions considered in the simulation study and the
description of their parameters have been provided in
Section~\ref{s:simulation}.   It remains to describe the non-random grid of
points on which the estimation error has been evaluated. Consider the
hypercube $(-q, q)^d$ where $q=6$ when $f(x)$ is a mixture with $0<\pi<1$ and
$q=3$ is $\pi=1$.  On the interval $[-q,q]$, select $N_0=\lceil N^{1/d}\rceil$
equally spaced points, where $\lceil{x}\rceil$ is the smallest integer larger
than or equal to $x$ and usually $N=2000$ was used.  The Cartesian product of
these coordinates for $d$ coordinated axes produces a grid of
$N_\mathrm{pts}=N_0^d$ evaluation points.

\noindent
The following pages provided a summary for each simulation run followed by 
an overall summary.
 
\clearpage 

\centerline{\Large Case No.\, 1 }
\begin{verbatim}
Niter : 2500 
n : 500 
d : 4 
m : 3 
qN : 6 
Npts : 1296 
dp1$xi : 2 2 2 2 
dp1$Omega : 
         [,1]     [,2]     [,3]     [,4]
[1,]  1.00000  0.60714 -0.04464 -0.33705
[2,]  0.60714  1.00000  0.60714 -0.04464
[3,] -0.04464  0.60714  1.00000  0.60714
[4,] -0.33705 -0.04464  0.60714  1.00000
dp1$alpha :  6  3 -6 -3 
dp1$nu : 5 
mix.p : 0.3333 
dp2$xi : -2 -2 -2 -2 
dp2$Omega : 
       [,1]   [,2]   [,3]   [,4]
[1,] 1.0000 0.7500 0.5625 0.4219
[2,] 0.7500 1.0000 0.7500 0.5625
[3,] 0.5625 0.7500 1.0000 0.7500
[4,] 0.4219 0.5625 0.7500 1.0000
dp2$alpha : -6 -3  6  3 
dp2$nu : 5 
\end{verbatim}
\begin{table}[ht]
\centering
\caption{Error quantiles for a fixed grid of points} 
\begin{tabular}{lllllll}
  \hline
 & 25\% & 50\% & 75\% & 90\% & 95\% & 99\% \\ 
  \hline
e1 & 2.9E-05 & 9.79E-05 & 0.000348 & 0.00115 & 0.0028 & 0.0344 \\ 
  e2 & 2.9E-05 & 9.79E-05 & 0.000348 & 0.00115 & 0.0028 & 0.034 \\ 
  e3 & 2.89E-05 & 9.79E-05 & 0.000348 & 0.00115 & 0.0028 & 0.0341 \\ 
  e4 & 2.89E-05 & 9.79E-05 & 0.00035 & 0.00118 & 0.00299 & 0.0371 \\ 
   \hline
\end{tabular}
\end{table}
\begin{table}[ht]
\centering
\caption{Error quantiles evaluating at the observed sample points} 
\begin{tabular}{lllllll}
  \hline
 & 25\% & 50\% & 75\% & 90\% & 95\% & 99\% \\ 
  \hline
e1 & 0.0274 & 0.06 & 0.114 & 0.183 & 0.232 & 0.636 \\ 
  e2 & 0.0272 & 0.0596 & 0.114 & 0.182 & 0.232 & 0.634 \\ 
  e3 & 0.0273 & 0.0597 & 0.114 & 0.182 & 0.232 & 0.634 \\ 
  e4 & 0.0272 & 0.0596 & 0.114 & 0.182 & 0.232 & 0.63 \\ 
   \hline
\end{tabular}
\end{table}
\clearpage\centerline{\Large Case No.\, 2 }
\begin{verbatim}
Niter : 2500 
n : 500 
d : 4 
m : 2 
qN : 6 
Npts : 1296 
dp1$xi : 2 2 2 2 
dp1$Omega : 
         [,1]     [,2]     [,3]     [,4]
[1,]  1.00000  0.60714 -0.04464 -0.33705
[2,]  0.60714  1.00000  0.60714 -0.04464
[3,] -0.04464  0.60714  1.00000  0.60714
[4,] -0.33705 -0.04464  0.60714  1.00000
dp1$alpha :  6  3 -6 -3 
dp1$nu : 5 
mix.p : 0.3333 
dp2$xi : -2 -2 -2 -2 
dp2$Omega : 
       [,1]   [,2]   [,3]   [,4]
[1,] 1.0000 0.7500 0.5625 0.4219
[2,] 0.7500 1.0000 0.7500 0.5625
[3,] 0.5625 0.7500 1.0000 0.7500
[4,] 0.4219 0.5625 0.7500 1.0000
dp2$alpha : -6 -3  6  3 
dp2$nu : 5 
\end{verbatim}
\begin{table}[ht]
\centering
\caption{Error quantiles for a fixed grid of points} 
\begin{tabular}{lllllll}
  \hline
 & 25\% & 50\% & 75\% & 90\% & 95\% & 99\% \\ 
  \hline
e1 & 2.89E-05 & 9.79E-05 & 0.000348 & 0.00115 & 0.00279 & 0.0346 \\ 
  e2 & 2.9E-05 & 9.79E-05 & 0.000348 & 0.00115 & 0.00272 & 0.0303 \\ 
  e3 & 2.9E-05 & 9.79E-05 & 0.000348 & 0.00115 & 0.00273 & 0.0306 \\ 
  e4 & 2.9E-05 & 9.98E-05 & 0.000371 & 0.00167 & 0.00581 & 0.0577 \\ 
   \hline
\end{tabular}
\end{table}
\begin{table}[ht]
\centering
\caption{Error quantiles evaluating at the observed sample points} 
\begin{tabular}{lllllll}
  \hline
 & 25\% & 50\% & 75\% & 90\% & 95\% & 99\% \\ 
  \hline
e1 & 0.0274 & 0.06 & 0.114 & 0.182 & 0.232 & 0.622 \\ 
  e2 & 0.0255 & 0.0563 & 0.108 & 0.175 & 0.22 & 0.497 \\ 
  e3 & 0.0261 & 0.0568 & 0.108 & 0.175 & 0.22 & 0.497 \\ 
  e4 & 0.0252 & 0.0561 & 0.109 & 0.176 & 0.221 & 0.471 \\ 
   \hline
\end{tabular}
\end{table}
\clearpage\centerline{\Large Case No.\, 3 }
\begin{verbatim}
Niter : 2500 
n : 500 
d : 4 
m : 3 
qN : 6 
Npts : 1296 
dp1$xi : 2 2 2 2 
dp1$Omega : 
         [,1]     [,2]     [,3]     [,4]
[1,]  1.00000  0.60714 -0.04464 -0.33705
[2,]  0.60714  1.00000  0.60714 -0.04464
[3,] -0.04464  0.60714  1.00000  0.60714
[4,] -0.33705 -0.04464  0.60714  1.00000
dp1$alpha :  6  3 -6 -3 
mix.p : 0.3333 
dp2$xi : -2 -2 -2 -2 
dp2$Omega : 
       [,1]   [,2]   [,3]   [,4]
[1,] 1.0000 0.7500 0.5625 0.4219
[2,] 0.7500 1.0000 0.7500 0.5625
[3,] 0.5625 0.7500 1.0000 0.7500
[4,] 0.4219 0.5625 0.7500 1.0000
dp2$alpha : -6 -3  6  3 
\end{verbatim}
\begin{table}[ht]
\centering
\caption{Error quantiles for a fixed grid of points} 
\begin{tabular}{lllllll}
  \hline
 & 25\% & 50\% & 75\% & 90\% & 95\% & 99\% \\ 
  \hline
e1 & 5.13E-22 & 3.77E-12 & 6.97E-05 & 3.57E+03 & 1.81E+13 & 2.11E+36 \\ 
  e2 & 5.12E-22 & 3.76E-12 & 6.94E-05 & 3.58E+03 & 1.8E+13 & 2.15E+36 \\ 
  e3 & 5.11E-22 & 3.76E-12 & 6.94E-05 & 3.56E+03 & 1.81E+13 & 1.88E+36 \\ 
  e4 & 5.11E-20 & 5.64E-11 & 0.000709 & 1.25E+06 & 1.19E+20 & 1.2E+51 \\ 
   \hline
\end{tabular}
\end{table}
\begin{table}[ht]
\centering
\caption{Error quantiles evaluating at the observed sample points} 
\begin{tabular}{lllllll}
  \hline
 & 25\% & 50\% & 75\% & 90\% & 95\% & 99\% \\ 
  \hline
e1 & 0.0267 & 0.0569 & 0.0986 & 0.141 & 0.167 & 0.233 \\ 
  e2 & 0.0266 & 0.0568 & 0.0985 & 0.141 & 0.167 & 0.233 \\ 
  e3 & 0.0266 & 0.0568 & 0.0985 & 0.141 & 0.167 & 0.233 \\ 
  e4 & 0.0266 & 0.0569 & 0.0986 & 0.141 & 0.167 & 0.233 \\ 
   \hline
\end{tabular}
\end{table}
\clearpage\centerline{\Large Case No.\, 4 }
\begin{verbatim}
Niter : 2500 
n : 500 
d : 4 
m : 2 
qN : 6 
Npts : 1296 
dp1$xi : 2 2 2 2 
dp1$Omega : 
         [,1]     [,2]     [,3]     [,4]
[1,]  1.00000  0.60714 -0.04464 -0.33705
[2,]  0.60714  1.00000  0.60714 -0.04464
[3,] -0.04464  0.60714  1.00000  0.60714
[4,] -0.33705 -0.04464  0.60714  1.00000
dp1$alpha :  6  3 -6 -3 
mix.p : 0.3333 
dp2$xi : -2 -2 -2 -2 
dp2$Omega : 
       [,1]   [,2]   [,3]   [,4]
[1,] 1.0000 0.7500 0.5625 0.4219
[2,] 0.7500 1.0000 0.7500 0.5625
[3,] 0.5625 0.7500 1.0000 0.7500
[4,] 0.4219 0.5625 0.7500 1.0000
dp2$alpha : -6 -3  6  3 
\end{verbatim}
\begin{table}[ht]
\centering
\caption{Error quantiles for a fixed grid of points} 
\begin{tabular}{lllllll}
  \hline
 & 25\% & 50\% & 75\% & 90\% & 95\% & 99\% \\ 
  \hline
kde & 5.14E-22 & 3.79E-12 & 7.13E-05 & 4.21E+03 & 3.22E+13 & 1.29E+36 \\ 
  wkde & 5.01E-22 & 3.68E-12 & 6.75E-05 & 4.02E+03 & 3.03E+13 & 1.25E+36 \\ 
  wkdeA & 5E-22 & 3.7E-12 & 6.84E-05 & 4.06E+03 & 2.92E+13 & 1.19E+36 \\ 
  fill+wkde & 4.55E-16 & 4.4E-08 & 0.277 & 8.34E+12 & 9.88E+27 & 1.56E+61 \\ 
   \hline
\end{tabular}
\end{table}
\begin{table}[ht]
\centering
\caption{Error quantiles evaluating at the observed sample points} 
\begin{tabular}{lllllll}
  \hline
 & 25\% & 50\% & 75\% & 90\% & 95\% & 99\% \\ 
  \hline
kde & 0.0266 & 0.057 & 0.0986 & 0.141 & 0.167 & 0.234 \\ 
  wkde & 0.0264 & 0.0565 & 0.098 & 0.14 & 0.166 & 0.232 \\ 
  wkdeA & 0.0263 & 0.0563 & 0.0976 & 0.14 & 0.165 & 0.23 \\ 
  fill+wkde & 0.0265 & 0.0568 & 0.0985 & 0.141 & 0.166 & 0.231 \\ 
   \hline
\end{tabular}
\end{table}
\clearpage\centerline{\Large Case No.\, 5 }
\begin{verbatim}
Niter : 2500 
n : 500 
d : 4 
m : 3 
qN : 6 
Npts : 1296 
dp1$xi : 2 2 2 2 
dp1$Omega : 
         [,1]     [,2]     [,3]     [,4]
[1,]  1.00000  0.60714 -0.04464 -0.33705
[2,]  0.60714  1.00000  0.60714 -0.04464
[3,] -0.04464  0.60714  1.00000  0.60714
[4,] -0.33705 -0.04464  0.60714  1.00000
dp1$alpha :  6  3 -6 -3 
dp1$nu : 2 
mix.p : 0.3333 
dp2$xi : -2 -2 -2 -2 
dp2$Omega : 
       [,1]   [,2]   [,3]   [,4]
[1,] 1.0000 0.7500 0.5625 0.4219
[2,] 0.7500 1.0000 0.7500 0.5625
[3,] 0.5625 0.7500 1.0000 0.7500
[4,] 0.4219 0.5625 0.7500 1.0000
dp2$alpha : -6 -3  6  3 
dp2$nu : 2 
\end{verbatim}
\begin{table}[ht]
\centering
\caption{Error quantiles for a fixed grid of points} 
\begin{tabular}{lllllll}
  \hline
 & 25\% & 50\% & 75\% & 90\% & 95\% & 99\% \\ 
  \hline
kde & 0.000202 & 0.000462 & 0.00115 & 0.00321 & 0.00813 & 0.0485 \\ 
  wkde & 0.000202 & 0.000462 & 0.00116 & 0.00322 & 0.00818 & 0.0486 \\ 
  wkdeA & 0.000202 & 0.000462 & 0.00116 & 0.00322 & 0.00818 & 0.0486 \\ 
  fill+wkde & 0.000202 & 0.000462 & 0.00115 & 0.00321 & 0.00816 & 0.0486 \\ 
   \hline
\end{tabular}
\end{table}
\begin{table}[ht]
\centering
\caption{Error quantiles evaluating at the observed sample points} 
\begin{tabular}{lllllll}
  \hline
 & 25\% & 50\% & 75\% & 90\% & 95\% & 99\% \\ 
  \hline
kde & 0.025 & 0.0586 & 0.136 & 0.246 & 0.343 & 2.63 \\ 
  wkde & 0.025 & 0.0587 & 0.136 & 0.247 & 0.345 & 2.67 \\ 
  wkdeA & 0.025 & 0.0587 & 0.136 & 0.246 & 0.345 & 2.66 \\ 
  fill+wkde & 0.0249 & 0.0585 & 0.136 & 0.246 & 0.343 & 2.62 \\ 
   \hline
\end{tabular}
\end{table}
\clearpage\centerline{\Large Case No.\, 6 }
\begin{verbatim}
Niter : 2500 
n : 500 
d : 4 
m : 2 
qN : 6 
Npts : 1296 
dp1$xi : 2 2 2 2 
dp1$Omega : 
         [,1]     [,2]     [,3]     [,4]
[1,]  1.00000  0.60714 -0.04464 -0.33705
[2,]  0.60714  1.00000  0.60714 -0.04464
[3,] -0.04464  0.60714  1.00000  0.60714
[4,] -0.33705 -0.04464  0.60714  1.00000
dp1$alpha :  6  3 -6 -3 
dp1$nu : 2 
mix.p : 0.3333 
dp2$xi : -2 -2 -2 -2 
dp2$Omega : 
       [,1]   [,2]   [,3]   [,4]
[1,] 1.0000 0.7500 0.5625 0.4219
[2,] 0.7500 1.0000 0.7500 0.5625
[3,] 0.5625 0.7500 1.0000 0.7500
[4,] 0.4219 0.5625 0.7500 1.0000
dp2$alpha : -6 -3  6  3 
dp2$nu : 2 
\end{verbatim}
\begin{table}[ht]
\centering
\caption{Error quantiles for a fixed grid of points} 
\begin{tabular}{lllllll}
  \hline
 & 25\% & 50\% & 75\% & 90\% & 95\% & 99\% \\ 
  \hline
kde & 0.000208 & 0.000481 & 0.00121 & 0.00331 & 0.00827 & 0.0487 \\ 
  wkde & 0.00021 & 0.000486 & 0.00122 & 0.00319 & 0.00689 & 0.0398 \\ 
  wkdeA & 0.00021 & 0.000486 & 0.00121 & 0.00319 & 0.00694 & 0.0396 \\ 
  fill+wkde & 0.000205 & 0.000478 & 0.00121 & 0.00329 & 0.00775 & 0.0443 \\ 
   \hline
\end{tabular}
\end{table}
\begin{table}[ht]
\centering
\caption{Error quantiles evaluating at the observed sample points} 
\begin{tabular}{lllllll}
  \hline
 & 25\% & 50\% & 75\% & 90\% & 95\% & 99\% \\ 
  \hline
kde & 0.025 & 0.0587 & 0.135 & 0.245 & 0.341 &  2.6 \\ 
  wkde & 0.0182 & 0.0494 & 0.125 & 0.234 & 0.323 & 2.95 \\ 
  wkdeA & 0.0188 & 0.0496 & 0.125 & 0.234 & 0.323 & 2.95 \\ 
  fill+wkde & 0.0181 & 0.0485 & 0.124 & 0.228 & 0.302 & 1.73 \\ 
   \hline
\end{tabular}
\end{table}
\clearpage\centerline{\Large Case No.\, 7 }
\begin{verbatim}
Niter : 2500 
n : 500 
d : 4 
m : 3 
qN : 6 
Npts : 1296 
dp1$xi : 2 2 2 2 
dp1$Omega : 
         [,1]     [,2]     [,3]     [,4]
[1,]  1.00000  0.60714 -0.04464 -0.33705
[2,]  0.60714  1.00000  0.60714 -0.04464
[3,] -0.04464  0.60714  1.00000  0.60714
[4,] -0.33705 -0.04464  0.60714  1.00000
dp1$alpha :  6  3 -6 -3 
dp1$nu : 2 
mix.p : 0.6667 
dp2$xi : -2 -2 -2 -2 
dp2$Omega : 
       [,1]   [,2]   [,3]   [,4]
[1,] 1.0000 0.7500 0.5625 0.4219
[2,] 0.7500 1.0000 0.7500 0.5625
[3,] 0.5625 0.7500 1.0000 0.7500
[4,] 0.4219 0.5625 0.7500 1.0000
dp2$alpha : -6 -3  6  3 
dp2$nu : 2 
\end{verbatim}
\begin{table}[ht]
\centering
\caption{Error quantiles for a fixed grid of points} 
\begin{tabular}{lllllll}
  \hline
 & 25\% & 50\% & 75\% & 90\% & 95\% & 99\% \\ 
  \hline
kde & 0.000182 & 0.000429 & 0.00112 & 0.00323 & 0.00812 & 0.0477 \\ 
  wkde & 0.000182 & 0.000429 & 0.00112 & 0.00324 & 0.00815 & 0.0479 \\ 
  wkdeA & 0.000182 & 0.000429 & 0.00112 & 0.00324 & 0.00815 & 0.0479 \\ 
  fill+wkde & 0.000182 & 0.000429 & 0.00112 & 0.00324 & 0.00814 & 0.0477 \\ 
   \hline
\end{tabular}
\end{table}
\begin{table}[ht]
\centering
\caption{Error quantiles evaluating at the observed sample points} 
\begin{tabular}{lllllll}
  \hline
 & 25\% & 50\% & 75\% & 90\% & 95\% & 99\% \\ 
  \hline
kde & 0.025 & 0.0614 & 0.141 & 0.254 & 0.355 & 2.65 \\ 
  wkde & 0.0251 & 0.0616 & 0.142 & 0.254 & 0.357 & 2.69 \\ 
  wkdeA & 0.0251 & 0.0616 & 0.142 & 0.254 & 0.357 & 2.69 \\ 
  fill+wkde & 0.025 & 0.0614 & 0.141 & 0.254 & 0.355 & 2.65 \\ 
   \hline
\end{tabular}
\end{table}
\clearpage\centerline{\Large Case No.\, 8 }
\begin{verbatim}
Niter : 2500 
n : 500 
d : 4 
m : 2 
qN : 6 
Npts : 1296 
dp1$xi : 2 2 2 2 
dp1$Omega : 
         [,1]     [,2]     [,3]     [,4]
[1,]  1.00000  0.60714 -0.04464 -0.33705
[2,]  0.60714  1.00000  0.60714 -0.04464
[3,] -0.04464  0.60714  1.00000  0.60714
[4,] -0.33705 -0.04464  0.60714  1.00000
dp1$alpha :  6  3 -6 -3 
dp1$nu : 2 
mix.p : 0.6667 
dp2$xi : -2 -2 -2 -2 
dp2$Omega : 
       [,1]   [,2]   [,3]   [,4]
[1,] 1.0000 0.7500 0.5625 0.4219
[2,] 0.7500 1.0000 0.7500 0.5625
[3,] 0.5625 0.7500 1.0000 0.7500
[4,] 0.4219 0.5625 0.7500 1.0000
dp2$alpha : -6 -3  6  3 
dp2$nu : 2 
\end{verbatim}
\begin{table}[ht]
\centering
\caption{Error quantiles for a fixed grid of points} 
\begin{tabular}{lllllll}
  \hline
 & 25\% & 50\% & 75\% & 90\% & 95\% & 99\% \\ 
  \hline
kde & 0.000182 & 0.000433 & 0.00113 & 0.00321 & 0.00806 & 0.0478 \\ 
  wkde & 0.000184 & 0.000439 & 0.00114 & 0.00304 & 0.00702 & 0.0392 \\ 
  wkdeA & 0.000184 & 0.000439 & 0.00114 & 0.00305 & 0.00707 & 0.039 \\ 
  fill+wkde & 0.000179 & 0.000426 & 0.00113 & 0.00324 & 0.00769 & 0.0417 \\ 
   \hline
\end{tabular}
\end{table}
\begin{table}[ht]
\centering
\caption{Error quantiles evaluating at the observed sample points} 
\begin{tabular}{lllllll}
  \hline
 & 25\% & 50\% & 75\% & 90\% & 95\% & 99\% \\ 
  \hline
kde & 0.0251 & 0.0616 & 0.141 & 0.254 & 0.355 & 2.73 \\ 
  wkde & 0.0211 & 0.0552 & 0.131 & 0.243 & 0.337 & 3.06 \\ 
  wkdeA & 0.0218 & 0.0554 & 0.131 & 0.243 & 0.338 & 3.07 \\ 
  fill+wkde & 0.0184 & 0.0514 & 0.13 & 0.235 & 0.314 & 1.71 \\ 
   \hline
\end{tabular}
\end{table}
\clearpage\centerline{\Large Case No.\, 9 }
\begin{verbatim}
Niter : 2500 
n : 500 
d : 4 
m : 3 
qN : 3 
Npts : 1296 
dp1$xi : 0 0 0 0 
dp1$Omega : 
     [,1] [,2] [,3] [,4]
[1,]    1    0    0    0
[2,]    0    1    0    0
[3,]    0    0    1    0
[4,]    0    0    0    1
dp1$alpha : 0 0 0 0 
dp1$nu : 2 
mix.p : 1 
\end{verbatim}
\begin{table}[ht]
\centering
\caption{Error quantiles for a fixed grid of points} 
\begin{tabular}{lllllll}
  \hline
 & 25\% & 50\% & 75\% & 90\% & 95\% & 99\% \\ 
  \hline
kde & 0.00304 & 0.00495 & 0.00858 & 0.0182 & 0.0268 & 0.0472 \\ 
  wkde & 0.00304 & 0.00496 & 0.00861 & 0.0183 & 0.027 & 0.0475 \\ 
  wkdeA & 0.00304 & 0.00496 & 0.00861 & 0.0183 & 0.027 & 0.0475 \\ 
  fill+wkde & 0.00304 & 0.00495 & 0.00857 & 0.0182 & 0.0268 & 0.0471 \\ 
   \hline
\end{tabular}
\end{table}
\begin{table}[ht]
\centering
\caption{Error quantiles evaluating at the observed sample points} 
\begin{tabular}{lllllll}
  \hline
 & 25\% & 50\% & 75\% & 90\% & 95\% & 99\% \\ 
  \hline
kde & 0.0158 & 0.0326 & 0.0707 & 0.185 & 0.506 & 5.55 \\ 
  wkde & 0.0158 & 0.0328 & 0.0712 & 0.186 & 0.509 & 5.57 \\ 
  wkdeA & 0.0158 & 0.0328 & 0.0713 & 0.187 & 0.509 & 5.57 \\ 
  fill+wkde & 0.0157 & 0.0325 & 0.0706 & 0.185 & 0.504 & 5.52 \\ 
   \hline
\end{tabular}
\end{table}
\clearpage\centerline{\Large Case No.\, 10 }
\begin{verbatim}
Niter : 2500 
n : 500 
d : 4 
m : 2 
qN : 3 
Npts : 1296 
dp1$xi : 0 0 0 0 
dp1$Omega : 
     [,1] [,2] [,3] [,4]
[1,]    1    0    0    0
[2,]    0    1    0    0
[3,]    0    0    1    0
[4,]    0    0    0    1
dp1$alpha : 0 0 0 0 
mix.p : 1 
\end{verbatim}
\begin{table}[ht]
\centering
\caption{Error quantiles for a fixed grid of points} 
\begin{tabular}{lllllll}
  \hline
 & 25\% & 50\% & 75\% & 90\% & 95\% & 99\% \\ 
  \hline
kde & 0.00071 & 0.00213 & 0.00624 & 0.0197 & 0.0349 & 0.0938 \\ 
  wkde & 0.000712 & 0.00215 & 0.00614 & 0.018 & 0.0306 & 0.0815 \\ 
  wkdeA & 0.000712 & 0.00213 & 0.00615 & 0.018 & 0.0308 & 0.0812 \\ 
  fill+wkde & 0.000976 & 0.00403 & 0.0142 & 0.0364 & 0.0585 & 0.125 \\ 
   \hline
\end{tabular}
\end{table}
\begin{table}[ht]
\centering
\caption{Error quantiles evaluating at the observed sample points} 
\begin{tabular}{lllllll}
  \hline
 & 25\% & 50\% & 75\% & 90\% & 95\% & 99\% \\ 
  \hline
kde & 0.0152 & 0.0334 & 0.0643 & 0.113 & 0.164 & 0.388 \\ 
  wkde & 0.0149 & 0.0329 & 0.0617 & 0.103 & 0.146 & 0.335 \\ 
  wkdeA & 0.0151 & 0.0327 & 0.0601 &  0.1 & 0.143 & 0.333 \\ 
  fill+wkde & 0.0108 & 0.0228 & 0.0406 & 0.0715 & 0.106 & 0.239 \\ 
   \hline
\end{tabular}
\end{table}
\clearpage\centerline{\Large Case No.\, 11 }
\begin{verbatim}
Niter : 2500 
n : 500 
d : 4 
m : 3 
qN : 3 
Npts : 1296 
dp1$xi : 0 0 0 0 
dp1$Omega : 
     [,1] [,2] [,3] [,4]
[1,]    1    0    0    0
[2,]    0    1    0    0
[3,]    0    0    1    0
[4,]    0    0    0    1
dp1$alpha : 0 0 0 0 
mix.p : 1 
\end{verbatim}
\begin{table}[ht]
\centering
\caption{Error quantiles for a fixed grid of points} 
\begin{tabular}{lllllll}
  \hline
 & 25\% & 50\% & 75\% & 90\% & 95\% & 99\% \\ 
  \hline
kde & 0.00071 & 0.00214 & 0.00625 & 0.0198 & 0.0349 & 0.0942 \\ 
  wkde & 0.000712 & 0.00218 & 0.00622 & 0.0195 & 0.0342 & 0.0937 \\ 
  wkdeA & 0.000711 & 0.00215 & 0.00623 & 0.0196 & 0.0344 & 0.0931 \\ 
  fill+wkde & 0.000646 & 0.00257 & 0.00921 & 0.026 & 0.0435 & 0.105 \\ 
   \hline
\end{tabular}
\end{table}
\begin{table}[ht]
\centering
\caption{Error quantiles evaluating at the observed sample points} 
\begin{tabular}{lllllll}
  \hline
 & 25\% & 50\% & 75\% & 90\% & 95\% & 99\% \\ 
  \hline
kde & 0.0152 & 0.0334 & 0.0643 & 0.113 & 0.164 & 0.389 \\ 
  wkde & 0.0153 & 0.0338 & 0.0651 & 0.114 & 0.164 & 0.387 \\ 
  wkdeA & 0.0153 & 0.0337 & 0.0648 & 0.113 & 0.162 & 0.382 \\ 
  fill+wkde & 0.0126 & 0.0275 & 0.0532 & 0.0979 & 0.143 & 0.336 \\ 
   \hline
\end{tabular}
\end{table}
\clearpage\centerline{\Large Case No.\, 12 }
\begin{verbatim}
Niter : 2500 
n : 500 
d : 4 
m : 3 
qN : 3 
Npts : 1296 
dp1$xi : 0 0 0 0 
dp1$Omega : 
     [,1] [,2] [,3] [,4]
[1,]    1    0    0    0
[2,]    0    1    0    0
[3,]    0    0    1    0
[4,]    0    0    0    1
dp1$alpha :  6  3 -6 -3 
mix.p : 1 
\end{verbatim}
\begin{table}[ht]
\centering
\caption{Error quantiles for a fixed grid of points} 
\begin{tabular}{lllllll}
  \hline
 & 25\% & 50\% & 75\% & 90\% & 95\% & 99\% \\ 
  \hline
kde & 0.00143 & 0.0127 & 3.02E+12 & 8.48E+53 & 3.35E+90 &  Inf \\ 
  wkde & 0.00144 & 0.0127 & 2.94E+12 & 7.96E+53 & 2.92E+90 &  Inf \\ 
  wkdeA & 0.00143 & 0.0127 & 3.05E+12 & 8.25E+53 & 3.1E+90 &  Inf \\ 
  fill+wkde & 0.00165 & 0.016 & 1.55E+14 & 1.08E+60 & 4.45E+100 &  Inf \\ 
   \hline
\end{tabular}
\end{table}
\begin{table}[ht]
\centering
\caption{Error quantiles evaluating at the observed sample points} 
\begin{tabular}{lllllll}
  \hline
 & 25\% & 50\% & 75\% & 90\% & 95\% & 99\% \\ 
  \hline
kde & 0.0198 & 0.0428 & 0.0796 & 0.14 & 0.206 & 0.49 \\ 
  wkde & 0.0197 & 0.0428 & 0.0805 & 0.141 & 0.207 & 0.488 \\ 
  wkdeA & 0.0198 & 0.043 & 0.0801 & 0.141 & 0.205 & 0.485 \\ 
  fill+wkde & 0.0177 & 0.0385 & 0.0718 & 0.126 & 0.187 & 0.439 \\ 
   \hline
\end{tabular}
\end{table}
\clearpage\centerline{\Large Case No.\, 13 }
\begin{verbatim}
Niter : 2500 
n : 500 
d : 4 
m : 2 
qN : 3 
Npts : 1296 
dp1$xi : 0 0 0 0 
dp1$Omega : 
     [,1] [,2] [,3] [,4]
[1,]    1    0    0    0
[2,]    0    1    0    0
[3,]    0    0    1    0
[4,]    0    0    0    1
dp1$alpha :  6  3 -6 -3 
mix.p : 1 
\end{verbatim}
\begin{table}[ht]
\centering
\caption{Error quantiles for a fixed grid of points} 
\begin{tabular}{lllllll}
  \hline
 & 25\% & 50\% & 75\% & 90\% & 95\% & 99\% \\ 
  \hline
kde & 0.00143 & 0.0127 & 2.91E+12 & 8.23E+53 & 3.23E+90 &  Inf \\ 
  wkde & 0.00141 & 0.0121 & 1.94E+12 & 5.96E+53 & 2.23E+90 &  Inf \\ 
  wkdeA & 0.00141 & 0.0121 & 2.17E+12 & 6.34E+53 & 2.42E+90 &  Inf \\ 
  fill+wkde & 0.00219 & 0.023 & 2.33E+16 & 1.55E+63 & 3.62E+105 &  Inf \\ 
   \hline
\end{tabular}
\end{table}
\begin{table}[ht]
\centering
\caption{Error quantiles evaluating at the observed sample points} 
\begin{tabular}{lllllll}
  \hline
 & 25\% & 50\% & 75\% & 90\% & 95\% & 99\% \\ 
  \hline
kde & 0.0197 & 0.0426 & 0.0794 & 0.14 & 0.205 & 0.487 \\ 
  wkde & 0.0189 & 0.0414 & 0.0778 & 0.135 & 0.195 & 0.458 \\ 
  wkdeA & 0.019 & 0.0413 & 0.0761 & 0.131 & 0.19 & 0.448 \\ 
  fill+wkde & 0.0157 & 0.0341 & 0.0621 & 0.0997 & 0.144 & 0.336 \\ 
   \hline
\end{tabular}
\end{table}
\clearpage\centerline{\Large Case No.\, 14 }
\begin{verbatim}
Niter : 2500 
n : 500 
d : 4 
m : 3 
qN : 3 
Npts : 1296 
dp1$xi : 0 0 0 0 
dp1$Omega : 
         [,1]     [,2]     [,3]     [,4]
[1,]  1.00000  0.60714 -0.04464 -0.33705
[2,]  0.60714  1.00000  0.60714 -0.04464
[3,] -0.04464  0.60714  1.00000  0.60714
[4,] -0.33705 -0.04464  0.60714  1.00000
dp1$alpha : 0 0 0 0 
mix.p : 1 
\end{verbatim}
\begin{table}[ht]
\centering
\caption{Error quantiles for a fixed grid of points} 
\begin{tabular}{lllllll}
  \hline
 & 25\% & 50\% & 75\% & 90\% & 95\% & 99\% \\ 
  \hline
kde & 2.01E-05 & 0.00195 & 0.033 & 0.283 & 1.31 &  122 \\ 
  wkde & 1.98E-05 & 0.00193 & 0.0325 & 0.269 &  1.2 &  104 \\ 
  wkdeA & 1.99E-05 & 0.00194 & 0.0327 & 0.273 & 1.23 &  107 \\ 
  fill+wkde & 8.63E-05 & 0.00794 & 0.429 & 5.43E+03 & 3.73E+06 & 1.02E+12 \\ 
   \hline
\end{tabular}
\end{table}
\begin{table}[ht]
\centering
\caption{Error quantiles evaluating at the observed sample points} 
\begin{tabular}{lllllll}
  \hline
 & 25\% & 50\% & 75\% & 90\% & 95\% & 99\% \\ 
  \hline
kde & 0.0265 & 0.0561 & 0.0977 & 0.158 & 0.228 &  0.5 \\ 
  wkde & 0.0265 & 0.0562 & 0.0977 & 0.157 & 0.227 & 0.498 \\ 
  wkdeA & 0.0265 & 0.0562 & 0.0978 & 0.157 & 0.227 & 0.498 \\ 
  fill+wkde & 0.0266 & 0.0562 & 0.0973 & 0.155 & 0.223 & 0.49 \\ 
   \hline
\end{tabular}
\end{table}
\clearpage\centerline{\Large Case No.\, 15 }
\begin{verbatim}
Niter : 2500 
n : 500 
d : 4 
m : 2 
qN : 3 
Npts : 1296 
dp1$xi : 0 0 0 0 
dp1$Omega : 
         [,1]     [,2]     [,3]     [,4]
[1,]  1.00000  0.60714 -0.04464 -0.33705
[2,]  0.60714  1.00000  0.60714 -0.04464
[3,] -0.04464  0.60714  1.00000  0.60714
[4,] -0.33705 -0.04464  0.60714  1.00000
dp1$alpha : 0 0 0 0 
mix.p : 1 
\end{verbatim}
\begin{table}[ht]
\centering
\caption{Error quantiles for a fixed grid of points} 
\begin{tabular}{lllllll}
  \hline
 & 25\% & 50\% & 75\% & 90\% & 95\% & 99\% \\ 
  \hline
kde & 2E-05 & 0.00194 & 0.0328 & 0.274 &  1.2 &  203 \\ 
  wkde & 1.55E-05 & 0.00159 & 0.0267 & 0.202 & 0.742 & 28.1 \\ 
  wkdeA & 1.66E-05 & 0.00169 & 0.0284 & 0.218 & 0.831 & 38.7 \\ 
  fill+wkde & 0.00882 & 0.391 & 9.27E+03 & 3.55E+10 & 1.78E+15 & 6.86E+25 \\ 
   \hline
\end{tabular}
\end{table}
\begin{table}[ht]
\centering
\caption{Error quantiles evaluating at the observed sample points} 
\begin{tabular}{lllllll}
  \hline
 & 25\% & 50\% & 75\% & 90\% & 95\% & 99\% \\ 
  \hline
kde & 0.0265 & 0.0561 & 0.0976 & 0.158 & 0.229 & 0.503 \\ 
  wkde & 0.0249 & 0.053 & 0.0928 & 0.149 & 0.215 & 0.472 \\ 
  wkdeA & 0.0249 & 0.0527 & 0.092 & 0.148 & 0.211 & 0.458 \\ 
  fill+wkde & 0.0257 & 0.0543 & 0.0926 & 0.138 & 0.191 & 0.423 \\ 
   \hline
\end{tabular}
\end{table}
\clearpage\centerline{\Large Case No.\, 16 }
\begin{verbatim}
Niter : 2500 
n : 500 
d : 4 
m : 3 
qN : 3 
Npts : 1296 
dp1$xi : 0 0 0 0 
dp1$Omega : 
       [,1]   [,2]   [,3]   [,4]
[1,] 1.0000 0.7500 0.5625 0.4219
[2,] 0.7500 1.0000 0.7500 0.5625
[3,] 0.5625 0.7500 1.0000 0.7500
[4,] 0.4219 0.5625 0.7500 1.0000
dp1$alpha :  6  3 -6 -3 
mix.p : 1 
\end{verbatim}
\begin{table}[ht]
\centering
\caption{Error quantiles for a fixed grid of points} 
\begin{tabular}{lllllll}
  \hline
 & 25\% & 50\% & 75\% & 90\% & 95\% & 99\% \\ 
  \hline
kde & 3.11E-07 & 0.000968 & 4.54E+07 & 3.54E+48 & 3.09E+82 &  Inf \\ 
  wkde & 3.1E-07 & 0.000953 & 4.43E+07 & 3.39E+48 & 3.02E+82 &  Inf \\ 
  wkdeA & 3.11E-07 & 0.00096 & 4.48E+07 & 3.48E+48 & 3.03E+82 &  Inf \\ 
  fill+wkde & 7.65E-07 & 0.00272 & 7.55E+08 & 1.14E+53 & 1.31E+88 &  Inf \\ 
   \hline
\end{tabular}
\end{table}
\begin{table}[ht]
\centering
\caption{Error quantiles evaluating at the observed sample points} 
\begin{tabular}{lllllll}
  \hline
 & 25\% & 50\% & 75\% & 90\% & 95\% & 99\% \\ 
  \hline
kde & 0.0312 & 0.0665 & 0.117 & 0.184 & 0.257 & 0.58 \\ 
  wkde & 0.0306 & 0.0653 & 0.115 & 0.182 & 0.256 & 0.58 \\ 
  wkdeA & 0.0306 & 0.0654 & 0.116 & 0.182 & 0.255 & 0.579 \\ 
  fill+wkde & 0.0306 & 0.0653 & 0.115 & 0.179 & 0.25 & 0.568 \\ 
   \hline
\end{tabular}
\end{table}
\clearpage\centerline{\Large Case No.\, 17 }
\begin{verbatim}
Niter : 2500 
n : 500 
d : 4 
m : 2 
qN : 3 
Npts : 1296 
dp1$xi : 0 0 0 0 
dp1$Omega : 
       [,1]   [,2]   [,3]   [,4]
[1,] 1.0000 0.7500 0.5625 0.4219
[2,] 0.7500 1.0000 0.7500 0.5625
[3,] 0.5625 0.7500 1.0000 0.7500
[4,] 0.4219 0.5625 0.7500 1.0000
dp1$alpha :  6  3 -6 -3 
mix.p : 1 
\end{verbatim}
\begin{table}[ht]
\centering
\caption{Error quantiles for a fixed grid of points} 
\begin{tabular}{lllllll}
  \hline
 & 25\% & 50\% & 75\% & 90\% & 95\% & 99\% \\ 
  \hline
kde & 3.1E-07 & 0.00096 & 4.36E+07 & 3.15E+48 & 1.37E+82 &  Inf \\ 
  wkde & 2.85E-07 & 0.000778 & 2.73E+07 & 1.89E+48 & 8.26E+81 &  Inf \\ 
  wkdeA & 2.9E-07 & 0.000807 & 3.02E+07 & 2.14E+48 & 8.73E+81 &  Inf \\ 
  fill+wkde & 0.000213 & 0.0343 & 9.94E+13 & 7.86E+59 & 3.93E+99 &  Inf \\ 
   \hline
\end{tabular}
\end{table}
\begin{table}[ht]
\centering
\caption{Error quantiles evaluating at the observed sample points} 
\begin{tabular}{lllllll}
  \hline
 & 25\% & 50\% & 75\% & 90\% & 95\% & 99\% \\ 
  \hline
kde & 0.0312 & 0.0665 & 0.117 & 0.184 & 0.257 & 0.574 \\ 
  wkde & 0.0274 & 0.0592 & 0.106 & 0.168 & 0.235 & 0.531 \\ 
  wkdeA & 0.0273 & 0.0588 & 0.105 & 0.165 & 0.229 & 0.518 \\ 
  fill+wkde & 0.0274 & 0.059 & 0.105 & 0.16 & 0.211 & 0.475 \\ 
   \hline
\end{tabular}
\end{table}
\clearpage\centerline{\Large Case No.\, 18 }
\begin{verbatim}
Niter : 2500 
n : 500 
d : 3 
m : 2 
qN : 3 
Npts : 1000 
dp1$xi : 0 0 0 
dp1$Omega : 
     [,1] [,2] [,3]
[1,] 1.00  0.8 0.32
[2,] 0.80  1.0 0.40
[3,] 0.32  0.4 1.00
dp1$alpha : 6 3 0 
mix.p : 1 
\end{verbatim}
\begin{table}[ht]
\centering
\caption{Error quantiles for a fixed grid of points} 
\begin{tabular}{lllllll}
  \hline
 & 25\% & 50\% & 75\% & 90\% & 95\% & 99\% \\ 
  \hline
kde & 5.5E-06 & 0.00257 & 0.0482 & 5.2E+03 & 3.79E+08 & 4.83E+14 \\ 
  wkde & 5.23E-06 & 0.00243 & 0.0436 & 5.26E+03 & 2.59E+08 & 3.08E+14 \\ 
  wkdeA & 5.32E-06 & 0.00244 & 0.0438 & 4.1E+03 & 2.66E+08 & 3.18E+14 \\ 
  fill+wkde & 2.73E-05 & 0.0091 & 0.149 & 6.56E+07 & 3.94E+13 & 1.33E+22 \\ 
   \hline
\end{tabular}
\end{table}
\begin{table}[ht]
\centering
\caption{Error quantiles evaluating at the observed sample points} 
\begin{tabular}{lllllll}
  \hline
 & 25\% & 50\% & 75\% & 90\% & 95\% & 99\% \\ 
  \hline
kde & 0.0256 & 0.0541 & 0.0941 & 0.144 & 0.189 & 0.357 \\ 
  wkde & 0.0238 & 0.0505 & 0.0882 & 0.135 & 0.177 & 0.335 \\ 
  wkdeA & 0.0234 & 0.0497 & 0.0866 & 0.133 & 0.173 & 0.327 \\ 
  fill+wkde & 0.0239 & 0.0504 & 0.0872 & 0.132 & 0.17 & 0.324 \\ 
   \hline
\end{tabular}
\end{table}
\clearpage\centerline{\Large Case No.\, 19 }
\begin{verbatim}
Niter : 2500 
n : 500 
d : 3 
m : 2 
qN : 3 
Npts : 1000 
dp1$xi : 0 0 0 
dp1$Omega : 
     [,1] [,2] [,3]
[1,] 1.00  0.8 0.32
[2,] 0.80  1.0 0.40
[3,] 0.32  0.4 1.00
dp1$alpha : 0 0 0 
mix.p : 1 
\end{verbatim}
\begin{table}[ht]
\centering
\caption{Error quantiles for a fixed grid of points} 
\begin{tabular}{lllllll}
  \hline
 & 25\% & 50\% & 75\% & 90\% & 95\% & 99\% \\ 
  \hline
kde & 8.01E-05 & 0.00341 & 0.0216 & 0.0572 & 0.0879 & 0.187 \\ 
  wkde & 7.23E-05 & 0.00298 & 0.0193 & 0.0501 & 0.0767 & 0.156 \\ 
  wkdeA & 7.29E-05 & 0.00303 & 0.0195 & 0.0501 & 0.0762 & 0.153 \\ 
  fill+wkde & 0.00148 & 0.0128 & 0.0468 & 0.146 & 0.475 & 33.2 \\ 
   \hline
\end{tabular}
\end{table}
\begin{table}[ht]
\centering
\caption{Error quantiles evaluating at the observed sample points} 
\begin{tabular}{lllllll}
  \hline
 & 25\% & 50\% & 75\% & 90\% & 95\% & 99\% \\ 
  \hline
kde & 0.0198 & 0.0415 & 0.0713 & 0.108 & 0.141 & 0.258 \\ 
  wkde & 0.0186 & 0.039 & 0.0669 & 0.101 & 0.13 & 0.237 \\ 
  wkdeA & 0.0184 & 0.0386 & 0.0662 & 0.0992 & 0.128 & 0.232 \\ 
  fill+wkde & 0.019 & 0.0397 & 0.0674 & 0.0991 & 0.126 & 0.232 \\ 
   \hline
\end{tabular}
\end{table}
\clearpage\centerline{\Large Case No.\, 20 }
\begin{verbatim}
Niter : 2500 
n : 500 
d : 3 
m : 2 
qN : 3 
Npts : 1000 
dp1$xi : 0 0 0 
dp1$Omega : 
     [,1] [,2] [,3]
[1,]  1.0  0.8 -0.1
[2,]  0.8  1.0  0.4
[3,] -0.1  0.4  1.0
dp1$alpha : 0 0 0 
mix.p : 1 
\end{verbatim}
\begin{table}[ht]
\centering
\caption{Error quantiles for a fixed grid of points} 
\begin{tabular}{lllllll}
  \hline
 & 25\% & 50\% & 75\% & 90\% & 95\% & 99\% \\ 
  \hline
kde & 3.96E-05 & 0.00577 & 0.0462 & 0.143 & 0.248 & 1.12 \\ 
  wkde & 3.3E-05 & 0.00534 & 0.0436 & 0.131 & 0.219 & 0.777 \\ 
  wkdeA & 3.47E-05 & 0.00544 & 0.0441 & 0.133 & 0.221 & 0.785 \\ 
  fill+wkde & 0.0117 & 0.105 & 23.3 & 1.39E+05 & 1.16E+08 & 2.87E+15 \\ 
   \hline
\end{tabular}
\end{table}
\begin{table}[ht]
\centering
\caption{Error quantiles evaluating at the observed sample points} 
\begin{tabular}{lllllll}
  \hline
 & 25\% & 50\% & 75\% & 90\% & 95\% & 99\% \\ 
  \hline
kde & 0.0308 & 0.0642 & 0.108 & 0.156 & 0.202 & 0.367 \\ 
  wkde & 0.0304 & 0.0634 & 0.106 & 0.154 & 0.197 & 0.347 \\ 
  wkdeA & 0.0303 & 0.0632 & 0.106 & 0.153 & 0.196 & 0.342 \\ 
  fill+wkde & 0.0311 & 0.0647 & 0.108 & 0.154 & 0.194 & 0.343 \\ 
   \hline
\end{tabular}
\end{table}
\clearpage\centerline{\Large Case No.\, 21 }
\begin{verbatim}
Niter : 2500 
n : 500 
d : 3 
m : 2 
qN : 3 
Npts : 1000 
dp1$xi : 0 0 0 
dp1$Omega : 
       [,1] [,2]   [,3]
[1,] 1.0000 0.75 0.5625
[2,] 0.7500 1.00 0.7500
[3,] 0.5625 0.75 1.0000
dp1$alpha : 0 0 0 
mix.p : 1 
\end{verbatim}
\begin{table}[ht]
\centering
\caption{Error quantiles for a fixed grid of points} 
\begin{tabular}{lllllll}
  \hline
 & 25\% & 50\% & 75\% & 90\% & 95\% & 99\% \\ 
  \hline
kde & 9E-06 & 0.00113 & 0.0174 & 0.0557 & 0.0893 & 0.194 \\ 
  wkde & 8.21E-06 & 0.000996 & 0.0153 & 0.0489 & 0.0784 & 0.164 \\ 
  wkdeA & 8.28E-06 & 0.00101 & 0.0155 & 0.049 & 0.078 & 0.161 \\ 
  fill+wkde & 0.000244 & 0.00794 & 0.0425 & 0.138 & 0.354 & 10.8 \\ 
   \hline
\end{tabular}
\end{table}
\begin{table}[ht]
\centering
\caption{Error quantiles evaluating at the observed sample points} 
\begin{tabular}{lllllll}
  \hline
 & 25\% & 50\% & 75\% & 90\% & 95\% & 99\% \\ 
  \hline
kde & 0.0224 & 0.047 & 0.0803 & 0.119 & 0.154 & 0.278 \\ 
  wkde & 0.0209 & 0.0439 & 0.0752 & 0.112 & 0.143 & 0.255 \\ 
  wkdeA & 0.0207 & 0.0435 & 0.0744 & 0.11 & 0.14 & 0.25 \\ 
  fill+wkde & 0.0214 & 0.0449 & 0.0761 & 0.111 & 0.139 & 0.25 \\ 
   \hline
\end{tabular}
\end{table}
\clearpage\centerline{\Large Case No.\, 22 }
\begin{verbatim}
Niter : 2500 
n : 500 
d : 3 
m : 2 
qN : 3 
Npts : 1000 
dp1$xi : 0 0 0 
dp1$Omega : 
     [,1] [,2] [,3]
[1,]    1    0    0
[2,]    0    1    0
[3,]    0    0    1
dp1$alpha : 0 0 0 
dp1$nu : 5 
mix.p : 1 
\end{verbatim}
\begin{table}[ht]
\centering
\caption{Error quantiles for a fixed grid of points} 
\begin{tabular}{lllllll}
  \hline
 & 25\% & 50\% & 75\% & 90\% & 95\% & 99\% \\ 
  \hline
kde & 0.00828 & 0.0143 & 0.0246 & 0.0434 & 0.0588 & 0.0937 \\ 
  wkde & 0.00872 & 0.0147 & 0.0239 & 0.0399 & 0.0559 &  0.1 \\ 
  wkdeA & 0.00853 & 0.0144 & 0.0235 & 0.0388 & 0.0538 & 0.0949 \\ 
  fill+wkde & 0.0081 & 0.014 & 0.0232 & 0.0366 & 0.05 & 0.0867 \\ 
   \hline
\end{tabular}
\end{table}
\begin{table}[ht]
\centering
\caption{Error quantiles evaluating at the observed sample points} 
\begin{tabular}{lllllll}
  \hline
 & 25\% & 50\% & 75\% & 90\% & 95\% & 99\% \\ 
  \hline
kde & 0.0157 & 0.0329 & 0.0573 & 0.0914 & 0.139 & 0.474 \\ 
  wkde & 0.0134 & 0.0288 & 0.0524 & 0.0858 & 0.125 & 0.454 \\ 
  wkdeA & 0.0133 & 0.0285 & 0.0513 & 0.0832 & 0.122 & 0.448 \\ 
  fill+wkde & 0.0107 & 0.0234 & 0.0434 & 0.07 & 0.0988 & 0.318 \\ 
   \hline
\end{tabular}
\end{table}
\clearpage\centerline{\Large Case No.\, 23 }
\begin{verbatim}
Niter : 2500 
n : 500 
d : 3 
m : 2 
qN : 3 
Npts : 1000 
dp1$xi : 0 0 0 
dp1$Omega : 
       [,1] [,2]   [,3]
[1,] 1.0000 0.75 0.5625
[2,] 0.7500 1.00 0.7500
[3,] 0.5625 0.75 1.0000
dp1$alpha : 0 0 0 
dp1$nu : 5 
mix.p : 1 
\end{verbatim}
\begin{table}[ht]
\centering
\caption{Error quantiles for a fixed grid of points} 
\begin{tabular}{lllllll}
  \hline
 & 25\% & 50\% & 75\% & 90\% & 95\% & 99\% \\ 
  \hline
kde & 0.00369 & 0.00857 & 0.0224 & 0.0532 & 0.0811 & 0.155 \\ 
  wkde & 0.00373 & 0.00845 & 0.0194 & 0.0419 & 0.0652 & 0.126 \\ 
  wkdeA & 0.00371 & 0.00834 & 0.0192 & 0.0415 & 0.064 & 0.122 \\ 
  fill+wkde & 0.00354 & 0.00836 & 0.0221 & 0.0496 & 0.075 & 0.144 \\ 
   \hline
\end{tabular}
\end{table}
\begin{table}[ht]
\centering
\caption{Error quantiles evaluating at the observed sample points} 
\begin{tabular}{lllllll}
  \hline
 & 25\% & 50\% & 75\% & 90\% & 95\% & 99\% \\ 
  \hline
kde & 0.0275 & 0.0576 & 0.098 & 0.144 & 0.197 & 0.639 \\ 
  wkde & 0.0229 & 0.0492 & 0.0863 & 0.131 & 0.181 & 0.61 \\ 
  wkdeA & 0.0232 & 0.0492 & 0.0854 & 0.129 & 0.178 & 0.607 \\ 
  fill+wkde & 0.024 & 0.0512 & 0.0894 & 0.132 & 0.167 & 0.517 \\ 
   \hline
\end{tabular}
\end{table}
\clearpage\centerline{\Large Case No.\, 24 }
\begin{verbatim}
Niter : 2500 
n : 500 
d : 3 
m : 2 
qN : 3 
Npts : 1000 
dp1$xi : 0 0 0 
dp1$Omega : 
       [,1] [,2]   [,3]
[1,] 1.0000 0.75 0.5625
[2,] 0.7500 1.00 0.7500
[3,] 0.5625 0.75 1.0000
dp1$alpha :  3  6 -6 
dp1$nu : 5 
mix.p : 1 
\end{verbatim}
\begin{table}[ht]
\centering
\caption{Error quantiles for a fixed grid of points} 
\begin{tabular}{lllllll}
  \hline
 & 25\% & 50\% & 75\% & 90\% & 95\% & 99\% \\ 
  \hline
kde & 0.000145 & 0.00468 & 0.0195 & 0.0651 & 0.126 & 0.422 \\ 
  wkde & 0.000145 & 0.00462 & 0.0184 & 0.0546 & 0.109 & 0.388 \\ 
  wkdeA & 0.000145 & 0.00461 & 0.0183 & 0.0543 & 0.108 & 0.386 \\ 
  fill+wkde & 0.000524 & 0.00638 & 0.027 & 0.101 & 0.264 & 1.78 \\ 
   \hline
\end{tabular}
\end{table}
\begin{table}[ht]
\centering
\caption{Error quantiles evaluating at the observed sample points} 
\begin{tabular}{lllllll}
  \hline
 & 25\% & 50\% & 75\% & 90\% & 95\% & 99\% \\ 
  \hline
kde & 0.0363 & 0.0771 & 0.137 & 0.21 & 0.274 & 0.81 \\ 
  wkde & 0.0311 & 0.0685 & 0.126 & 0.198 & 0.26 & 0.784 \\ 
  wkdeA & 0.0311 & 0.0681 & 0.125 & 0.196 & 0.258 & 0.782 \\ 
  fill+wkde & 0.0318 & 0.0702 & 0.131 & 0.205 & 0.256 & 0.689 \\ 
   \hline
\end{tabular}
\end{table}
\clearpage\centerline{\Large Case No.\, 25 }
\begin{verbatim}
Niter : 2500 
n : 500 
d : 5 
m : 2 
qN : 3 
Npts : 3125 
dp1$xi : 0 0 0 0 0 
dp1$Omega : 
       [,1]   [,2]   [,3]   [,4]   [,5]
[1,] 1.0000 0.7500 0.5625 0.4219 0.3164
[2,] 0.7500 1.0000 0.7500 0.5625 0.4219
[3,] 0.5625 0.7500 1.0000 0.7500 0.5625
[4,] 0.4219 0.5625 0.7500 1.0000 0.7500
[5,] 0.3164 0.4219 0.5625 0.7500 1.0000
dp1$alpha :  6  3  0 -6 -3 
mix.p : 1 
\end{verbatim}
\begin{table}[ht]
\centering
\caption{Error quantiles for a fixed grid of points} 
\begin{tabular}{lllllll}
  \hline
 & 25\% & 50\% & 75\% & 90\% & 95\% & 99\% \\ 
  \hline
kde & 4.95E-10 & 1.9E-05 & 1.29E+07 & 1.84E+53 & 2.87E+84 &  Inf \\ 
  wkde & 4.16E-10 & 1.39E-05 & 6.14E+06 & 5.23E+52 & 1.83E+84 &  Inf \\ 
  wkdeA & 4.35E-10 & 1.52E-05 & 7.75E+06 & 8.53E+52 & 2.11E+84 &  Inf \\ 
  fill+wkde & 0.000446 & 0.143 & 6.02E+15 & 2.72E+70 & 5.5E+102 &  Inf \\ 
   \hline
\end{tabular}
\end{table}
\begin{table}[ht]
\centering
\caption{Error quantiles evaluating at the observed sample points} 
\begin{tabular}{lllllll}
  \hline
 & 25\% & 50\% & 75\% & 90\% & 95\% & 99\% \\ 
  \hline
kde & 0.031 & 0.0671 & 0.125 & 0.23 & 0.353 & 0.901 \\ 
  wkde & 0.0253 & 0.0561 & 0.108 & 0.205 & 0.32 & 0.837 \\ 
  wkdeA & 0.0262 & 0.0573 & 0.109 & 0.204 & 0.317 & 0.833 \\ 
  fill+wkde & 0.0238 & 0.052 & 0.0959 & 0.16 & 0.245 & 0.65 \\ 
   \hline
\end{tabular}
\end{table}
\clearpage\centerline{\Large Case No.\, 26 }
\begin{verbatim}
Niter : 2500 
n : 500 
d : 5 
m : 3 
qN : 3 
Npts : 3125 
dp1$xi : 0 0 0 0 0 
dp1$Omega : 
       [,1]   [,2]   [,3]   [,4]   [,5]
[1,] 1.0000 0.7500 0.5625 0.4219 0.3164
[2,] 0.7500 1.0000 0.7500 0.5625 0.4219
[3,] 0.5625 0.7500 1.0000 0.7500 0.5625
[4,] 0.4219 0.5625 0.7500 1.0000 0.7500
[5,] 0.3164 0.4219 0.5625 0.7500 1.0000
dp1$alpha :  6  3  0 -6 -3 
mix.p : 1 
\end{verbatim}
\begin{table}[ht]
\centering
\caption{Error quantiles for a fixed grid of points} 
\begin{tabular}{lllllll}
  \hline
 & 25\% & 50\% & 75\% & 90\% & 95\% & 99\% \\ 
  \hline
kde & 4.73E-10 & 1.78E-05 & 9.03E+06 & 6.8E+52 & 1.45E+84 &  Inf \\ 
  wkde & 4.6E-10 & 1.69E-05 & 8.12E+06 & 6.19E+52 & 1.27E+84 &  Inf \\ 
  wkdeA & 4.66E-10 & 1.74E-05 & 8.65E+06 & 6.6E+52 & 1.38E+84 &  Inf \\ 
  fill+wkde & 8.95E-08 & 0.000937 & 5.11E+11 & 8.65E+60 & 1.23E+95 &  Inf \\ 
   \hline
\end{tabular}
\end{table}
\begin{table}[ht]
\centering
\caption{Error quantiles evaluating at the observed sample points} 
\begin{tabular}{lllllll}
  \hline
 & 25\% & 50\% & 75\% & 90\% & 95\% & 99\% \\ 
  \hline
kde & 0.031 & 0.0672 & 0.125 & 0.231 & 0.353 & 0.898 \\ 
  wkde & 0.0296 & 0.0647 & 0.122 & 0.229 & 0.352 & 0.91 \\ 
  wkdeA & 0.0302 & 0.0657 & 0.123 & 0.229 & 0.353 & 0.907 \\ 
  fill+wkde & 0.0286 & 0.062 & 0.115 & 0.208 & 0.323 & 0.834 \\ 
   \hline
\end{tabular}
\end{table}
\clearpage\centerline{\Large Case No.\, 27 }
\begin{verbatim}
Niter : 2500 
n : 500 
d : 5 
m : 2 
qN : 6 
Npts : 3125 
dp1$xi : 2 2 2 2 2 
dp1$Omega : 
         [,1]     [,2]     [,3]     [,4]     [,5]
[1,]  1.00000  0.60714 -0.04464 -0.33705 -0.23047
[2,]  0.60714  1.00000  0.60714 -0.04464 -0.33705
[3,] -0.04464  0.60714  1.00000  0.60714 -0.04464
[4,] -0.33705 -0.04464  0.60714  1.00000  0.60714
[5,] -0.23047 -0.33705 -0.04464  0.60714  1.00000
dp1$alpha :  6  3  0 -6 -3 
mix.p : 0.6667 
dp2$xi : -2 -2 -2 -2 -2 
dp2$Omega : 
       [,1]   [,2]   [,3]   [,4]   [,5]
[1,] 1.0000 0.7500 0.5625 0.4219 0.3164
[2,] 0.7500 1.0000 0.7500 0.5625 0.4219
[3,] 0.5625 0.7500 1.0000 0.7500 0.5625
[4,] 0.4219 0.5625 0.7500 1.0000 0.7500
[5,] 0.3164 0.4219 0.5625 0.7500 1.0000
dp2$alpha : -6 -3  0  6  3 
dp2$nu : 2 
\end{verbatim}
\begin{table}[ht]
\centering
\caption{Error quantiles for a fixed grid of points} 
\begin{tabular}{lllllll}
  \hline
 & 25\% & 50\% & 75\% & 90\% & 95\% & 99\% \\ 
  \hline
kde & 3.31E-06 & 3.14E-05 & 0.000159 & 0.00115 & 0.00416 & 0.0373 \\ 
  wkde & 3.72E-06 & 3.37E-05 & 0.000163 & 0.00113 & 0.00361 & 0.0272 \\ 
  wkdeA & 3.7E-06 & 3.39E-05 & 0.000163 & 0.00113 & 0.00359 & 0.0273 \\ 
  fill+wkde & 3.4E-06 & 3.24E-05 & 0.000162 & 0.0012 & 0.0042 & 0.0742 \\ 
   \hline
\end{tabular}
\end{table}
\begin{table}[ht]
\centering
\caption{Error quantiles evaluating at the observed sample points} 
\begin{tabular}{lllllll}
  \hline
 & 25\% & 50\% & 75\% & 90\% & 95\% & 99\% \\ 
  \hline
kde & 0.0274 & 0.0619 & 0.115 & 0.176 & 0.249 & 3.42 \\ 
  wkde & 0.0288 & 0.0638 & 0.116 & 0.174 & 0.223 & 3.57 \\ 
  wkdeA & 0.0295 & 0.0643 & 0.116 & 0.174 & 0.226 & 3.66 \\ 
  fill+wkde & 0.026 & 0.0619 & 0.115 & 0.17 & 0.208 & 1.74 \\ 
   \hline
\end{tabular}
\end{table}
\clearpage\centerline{\Large Case No.\, 28 }
\begin{verbatim}
Niter : 2500 
n : 500 
d : 5 
m : 3 
qN : 6 
Npts : 3125 
dp1$xi : 2 2 2 2 2 
dp1$Omega : 
         [,1]     [,2]     [,3]     [,4]     [,5]
[1,]  1.00000  0.60714 -0.04464 -0.33705 -0.23047
[2,]  0.60714  1.00000  0.60714 -0.04464 -0.33705
[3,] -0.04464  0.60714  1.00000  0.60714 -0.04464
[4,] -0.33705 -0.04464  0.60714  1.00000  0.60714
[5,] -0.23047 -0.33705 -0.04464  0.60714  1.00000
dp1$alpha :  6  3  0 -6 -3 
mix.p : 0.6667 
dp2$xi : -2 -2 -2 -2 -2 
dp2$Omega : 
       [,1]   [,2]   [,3]   [,4]   [,5]
[1,] 1.0000 0.7500 0.5625 0.4219 0.3164
[2,] 0.7500 1.0000 0.7500 0.5625 0.4219
[3,] 0.5625 0.7500 1.0000 0.7500 0.5625
[4,] 0.4219 0.5625 0.7500 1.0000 0.7500
[5,] 0.3164 0.4219 0.5625 0.7500 1.0000
dp2$alpha : -6 -3  0  6  3 
dp2$nu : 2 
\end{verbatim}
\begin{table}[ht]
\centering
\caption{Error quantiles for a fixed grid of points} 
\begin{tabular}{lllllll}
  \hline
 & 25\% & 50\% & 75\% & 90\% & 95\% & 99\% \\ 
  \hline
kde & 3.44E-06 & 3.15E-05 & 0.000149 & 0.00099 & 0.00317 & 0.0243 \\ 
  wkde & 3.43E-06 & 3.15E-05 & 0.000149 & 0.000987 & 0.00316 & 0.0245 \\ 
  wkdeA & 3.43E-06 & 3.15E-05 & 0.000149 & 0.000987 & 0.00316 & 0.0245 \\ 
  fill+wkde & 3.47E-06 & 3.19E-05 & 0.00015 & 0.000996 & 0.00319 & 0.0269 \\ 
   \hline
\end{tabular}
\end{table}
\begin{table}[ht]
\centering
\caption{Error quantiles evaluating at the observed sample points} 
\begin{tabular}{lllllll}
  \hline
 & 25\% & 50\% & 75\% & 90\% & 95\% & 99\% \\ 
  \hline
kde & 0.0275 & 0.0621 & 0.115 & 0.176 & 0.25 & 3.34 \\ 
  wkde & 0.0277 & 0.0625 & 0.116 & 0.177 & 0.253 & 3.46 \\ 
  wkdeA & 0.0277 & 0.0625 & 0.116 & 0.177 & 0.253 & 3.46 \\ 
  fill+wkde & 0.0275 & 0.0624 & 0.116 & 0.177 & 0.247 & 3.32 \\ 
   \hline
\end{tabular}
\end{table}
\clearpage\centerline{\Large Case No.\, 29 }
\begin{verbatim}
Niter : 2500 
n : 500 
d : 5 
m : 2 
qN : 6 
Npts : 7776 
dp1$xi : 2 2 2 2 2 
dp1$Omega : 
         [,1]     [,2]     [,3]     [,4]     [,5]
[1,]  1.00000  0.60714 -0.04464 -0.33705 -0.23047
[2,]  0.60714  1.00000  0.60714 -0.04464 -0.33705
[3,] -0.04464  0.60714  1.00000  0.60714 -0.04464
[4,] -0.33705 -0.04464  0.60714  1.00000  0.60714
[5,] -0.23047 -0.33705 -0.04464  0.60714  1.00000
dp1$alpha :  6  3  0 -6 -3 
dp1$nu : 5 
mix.p : 0.6667 
dp2$xi : -2 -2 -2 -2 -2 
dp2$Omega : 
       [,1]   [,2]   [,3]   [,4]   [,5]
[1,] 1.0000 0.7500 0.5625 0.4219 0.3164
[2,] 0.7500 1.0000 0.7500 0.5625 0.4219
[3,] 0.5625 0.7500 1.0000 0.7500 0.5625
[4,] 0.4219 0.5625 0.7500 1.0000 0.7500
[5,] 0.3164 0.4219 0.5625 0.7500 1.0000
dp2$alpha : -6 -3  0  6  3 
dp2$nu : 5 
\end{verbatim}
\begin{table}[ht]
\centering
\caption{Error quantiles for a fixed grid of points} 
\begin{tabular}{lllllll}
  \hline
 & 25\% & 50\% & 75\% & 90\% & 95\% & 99\% \\ 
  \hline
kde & 3.61E-06 & 1.26E-05 & 4.88E-05 & 0.000173 & 0.000381 & 0.00405 \\ 
  wkde & 3.61E-06 & 1.26E-05 & 4.88E-05 & 0.000173 & 0.000376 & 0.00349 \\ 
  wkdeA & 3.61E-06 & 1.26E-05 & 4.88E-05 & 0.000173 & 0.000377 & 0.00359 \\ 
  fill+wkde & 3.68E-06 & 1.32E-05 & 5.59E-05 & 0.000281 & 0.00115 & 0.0201 \\ 
   \hline
\end{tabular}
\end{table}
\begin{table}[ht]
\centering
\caption{Error quantiles evaluating at the observed sample points} 
\begin{tabular}{lllllll}
  \hline
 & 25\% & 50\% & 75\% & 90\% & 95\% & 99\% \\ 
  \hline
kde & 0.0231 & 0.0531 & 0.107 & 0.184 & 0.248 & 0.889 \\ 
  wkde & 0.0215 & 0.05 & 0.102 & 0.174 & 0.23 & 0.616 \\ 
  wkdeA & 0.0222 & 0.0506 & 0.102 & 0.174 & 0.231 & 0.629 \\ 
  fill+wkde & 0.021 & 0.0494 & 0.102 & 0.173 & 0.228 & 0.548 \\ 
   \hline
\end{tabular}
\end{table}
\clearpage\centerline{\Large Case No.\, 30 }
\begin{verbatim}
Niter : 2500 
n : 500 
d : 5 
m : 3 
qN : 6 
Npts : 7776 
dp1$xi : 2 2 2 2 2 
dp1$Omega : 
         [,1]     [,2]     [,3]     [,4]     [,5]
[1,]  1.00000  0.60714 -0.04464 -0.33705 -0.23047
[2,]  0.60714  1.00000  0.60714 -0.04464 -0.33705
[3,] -0.04464  0.60714  1.00000  0.60714 -0.04464
[4,] -0.33705 -0.04464  0.60714  1.00000  0.60714
[5,] -0.23047 -0.33705 -0.04464  0.60714  1.00000
dp1$alpha :  6  3  0 -6 -3 
dp1$nu : 5 
mix.p : 0.6667 
dp2$xi : -2 -2 -2 -2 -2 
dp2$Omega : 
       [,1]   [,2]   [,3]   [,4]   [,5]
[1,] 1.0000 0.7500 0.5625 0.4219 0.3164
[2,] 0.7500 1.0000 0.7500 0.5625 0.4219
[3,] 0.5625 0.7500 1.0000 0.7500 0.5625
[4,] 0.4219 0.5625 0.7500 1.0000 0.7500
[5,] 0.3164 0.4219 0.5625 0.7500 1.0000
dp2$alpha : -6 -3  0  6  3 
dp2$nu : 5 
\end{verbatim}
\begin{table}[ht]
\centering
\caption{Error quantiles for a fixed grid of points} 
\begin{tabular}{lllllll}
  \hline
 & 25\% & 50\% & 75\% & 90\% & 95\% & 99\% \\ 
  \hline
kde & 3.61E-06 & 1.26E-05 & 4.88E-05 & 0.000173 & 0.000381 & 0.00403 \\ 
  wkde & 3.61E-06 & 1.26E-05 & 4.88E-05 & 0.000173 & 0.000381 & 0.004 \\ 
  wkdeA & 3.61E-06 & 1.26E-05 & 4.88E-05 & 0.000173 & 0.000381 & 0.00401 \\ 
  fill+wkde & 3.61E-06 & 1.27E-05 & 5E-05 & 0.00019 & 0.000471 & 0.00742 \\ 
   \hline
\end{tabular}
\end{table}
\begin{table}[ht]
\centering
\caption{Error quantiles evaluating at the observed sample points} 
\begin{tabular}{lllllll}
  \hline
 & 25\% & 50\% & 75\% & 90\% & 95\% & 99\% \\ 
  \hline
kde & 0.0232 & 0.0532 & 0.108 & 0.184 & 0.248 & 0.869 \\ 
  wkde & 0.0231 & 0.0529 & 0.107 & 0.183 & 0.247 & 0.865 \\ 
  wkdeA & 0.0232 & 0.0531 & 0.107 & 0.183 & 0.247 & 0.868 \\ 
  fill+wkde & 0.0229 & 0.0528 & 0.107 & 0.183 & 0.246 & 0.844 \\ 
   \hline
\end{tabular}
\end{table}
\clearpage\centerline{\Large Case No.\, 31 }
\begin{verbatim}
Niter : 2500 
n : 500 
d : 5 
m : 2 
qN : 6 
Npts : 7776 
dp1$xi : 2 2 2 2 2 
dp1$Omega : 
         [,1]     [,2]     [,3]     [,4]     [,5]
[1,]  1.00000  0.60714 -0.04464 -0.33705 -0.23047
[2,]  0.60714  1.00000  0.60714 -0.04464 -0.33705
[3,] -0.04464  0.60714  1.00000  0.60714 -0.04464
[4,] -0.33705 -0.04464  0.60714  1.00000  0.60714
[5,] -0.23047 -0.33705 -0.04464  0.60714  1.00000
dp1$alpha :  6  3  0 -6 -3 
mix.p : 0.6667 
dp2$xi : -2 -2 -2 -2 -2 
dp2$Omega : 
       [,1]   [,2]   [,3]   [,4]   [,5]
[1,] 1.0000 0.7500 0.5625 0.4219 0.3164
[2,] 0.7500 1.0000 0.7500 0.5625 0.4219
[3,] 0.5625 0.7500 1.0000 0.7500 0.5625
[4,] 0.4219 0.5625 0.7500 1.0000 0.7500
[5,] 0.3164 0.4219 0.5625 0.7500 1.0000
dp2$alpha : -6 -3  0  6  3 
\end{verbatim}
\begin{table}[ht]
\centering
\caption{Error quantiles for a fixed grid of points} 
\begin{tabular}{lllllll}
  \hline
 & 25\% & 50\% & 75\% & 90\% & 95\% & 99\% \\ 
  \hline
kde & 3.83E-21 & 1.02E-11 & 0.00701 & 5.73E+18 & 8.16E+41 &  Inf \\ 
  wkde & 3.54E-21 & 9.36E-12 & 0.0064 & 5.27E+18 & 6.29E+41 &  Inf \\ 
  wkdeA & 3.69E-21 & 9.87E-12 & 0.0067 & 5.8E+18 & 1E+42 &  Inf \\ 
  fill+wkde & 5.42E-11 & 0.0174 & 1.67E+11 & 1.76E+37 & 1.27E+65 &  Inf \\ 
   \hline
\end{tabular}
\end{table}
\begin{table}[ht]
\centering
\caption{Error quantiles evaluating at the observed sample points} 
\begin{tabular}{lllllll}
  \hline
 & 25\% & 50\% & 75\% & 90\% & 95\% & 99\% \\ 
  \hline
kde & 0.0229 & 0.0501 & 0.0902 & 0.135 & 0.164 & 0.226 \\ 
  wkde & 0.0226 & 0.0497 & 0.0896 & 0.135 & 0.163 & 0.222 \\ 
  wkdeA & 0.0228 & 0.05 & 0.0901 & 0.135 & 0.164 & 0.227 \\ 
  fill+wkde & 0.0227 & 0.05 & 0.0901 & 0.135 & 0.164 & 0.221 \\ 
   \hline
\end{tabular}
\end{table}
\clearpage\centerline{\Large Case No.\, 32 }
\begin{verbatim}
Niter : 2500 
n : 500 
d : 5 
m : 3 
qN : 6 
Npts : 7776 
dp1$xi : 2 2 2 2 2 
dp1$Omega : 
         [,1]     [,2]     [,3]     [,4]     [,5]
[1,]  1.00000  0.60714 -0.04464 -0.33705 -0.23047
[2,]  0.60714  1.00000  0.60714 -0.04464 -0.33705
[3,] -0.04464  0.60714  1.00000  0.60714 -0.04464
[4,] -0.33705 -0.04464  0.60714  1.00000  0.60714
[5,] -0.23047 -0.33705 -0.04464  0.60714  1.00000
dp1$alpha :  6  3  0 -6 -3 
mix.p : 0.6667 
dp2$xi : -2 -2 -2 -2 -2 
dp2$Omega : 
       [,1]   [,2]   [,3]   [,4]   [,5]
[1,] 1.0000 0.7500 0.5625 0.4219 0.3164
[2,] 0.7500 1.0000 0.7500 0.5625 0.4219
[3,] 0.5625 0.7500 1.0000 0.7500 0.5625
[4,] 0.4219 0.5625 0.7500 1.0000 0.7500
[5,] 0.3164 0.4219 0.5625 0.7500 1.0000
dp2$alpha : -6 -3  0  6  3 
\end{verbatim}
\begin{table}[ht]
\centering
\caption{Error quantiles for a fixed grid of points} 
\begin{tabular}{lllllll}
  \hline
 & 25\% & 50\% & 75\% & 90\% & 95\% & 99\% \\ 
  \hline
kde & 2.94E-21 & 9.36E-12 & 0.00714 & 4.98E+18 & 5.1E+41 &  Inf \\ 
  wkde & 2.89E-21 & 9.29E-12 & 0.00707 & 4.93E+18 & 4.93E+41 &  Inf \\ 
  wkdeA & 2.93E-21 & 9.33E-12 & 0.00711 & 4.89E+18 & 4.98E+41 &  Inf \\ 
  fill+wkde & 3.05E-16 & 5.17E-07 & 1.99E+04 & 1.01E+28 & 1.47E+55 &  Inf \\ 
   \hline
\end{tabular}
\end{table}
\begin{table}[ht]
\centering
\caption{Error quantiles evaluating at the observed sample points} 
\begin{tabular}{lllllll}
  \hline
 & 25\% & 50\% & 75\% & 90\% & 95\% & 99\% \\ 
  \hline
kde & 0.0229 & 0.0501 & 0.0903 & 0.135 & 0.164 & 0.227 \\ 
  wkde & 0.0229 & 0.0501 & 0.0903 & 0.135 & 0.164 & 0.227 \\ 
  wkdeA & 0.0229 & 0.05 & 0.0903 & 0.135 & 0.164 & 0.227 \\ 
  fill+wkde & 0.0229 & 0.0501 & 0.0904 & 0.136 & 0.164 & 0.226 \\ 
   \hline
\end{tabular}
\end{table}
\clearpage\centerline{\Large Case No.\, 33 }
\begin{verbatim}
Niter : 2500 
n : 1000 
d : 5 
m : 3 
qN : 3 
Npts : 3125 
dp1$xi : 0 0 0 0 0 
dp1$Omega : 
       [,1]   [,2]   [,3]   [,4]   [,5]
[1,] 1.0000 0.7500 0.5625 0.4219 0.3164
[2,] 0.7500 1.0000 0.7500 0.5625 0.4219
[3,] 0.5625 0.7500 1.0000 0.7500 0.5625
[4,] 0.4219 0.5625 0.7500 1.0000 0.7500
[5,] 0.3164 0.4219 0.5625 0.7500 1.0000
dp1$alpha :  6  3  0 -6 -3 
mix.p : 1 
\end{verbatim}
\begin{table}[ht]
\centering
\caption{Error quantiles for a fixed grid of points} 
\begin{tabular}{lllllll}
  \hline
 & 25\% & 50\% & 75\% & 90\% & 95\% & 99\% \\ 
  \hline
kde & 1.79E-10 & 5.7E-06 & 1.02E+04 & 3.89E+46 & 3.37E+78 &  Inf \\ 
  wkde & 1.73E-10 & 5.59E-06 & 9.04E+03 & 3.35E+46 & 3.01E+78 &  Inf \\ 
  wkdeA & 1.76E-10 & 5.64E-06 & 9.62E+03 & 3.51E+46 & 3.11E+78 &  Inf \\ 
  fill+wkde & 1.37E-08 & 0.000521 & 7.61E+09 & 2.95E+56 & 2.3E+92 &  Inf \\ 
   \hline
\end{tabular}
\end{table}
\begin{table}[ht]
\centering
\caption{Error quantiles evaluating at the observed sample points} 
\begin{tabular}{lllllll}
  \hline
 & 25\% & 50\% & 75\% & 90\% & 95\% & 99\% \\ 
  \hline
kde & 0.0267 & 0.0578 & 0.106 & 0.188 & 0.287 & 0.728 \\ 
  wkde & 0.0245 & 0.0536 & 0.101 & 0.185 & 0.285 & 0.744 \\ 
  wkdeA & 0.0251 & 0.0545 & 0.102 & 0.184 & 0.284 & 0.738 \\ 
  fill+wkde & 0.0236 & 0.0514 & 0.0944 & 0.162 & 0.251 & 0.657 \\ 
   \hline
\end{tabular}
\end{table}
\clearpage\centerline{\Large Case No.\, 34 }
\begin{verbatim}
Niter : 2500 
n : 250 
d : 5 
m : 3 
qN : 3 
Npts : 3125 
dp1$xi : 0 0 0 0 0 
dp1$Omega : 
       [,1]   [,2]   [,3]   [,4]   [,5]
[1,] 1.0000 0.7500 0.5625 0.4219 0.3164
[2,] 0.7500 1.0000 0.7500 0.5625 0.4219
[3,] 0.5625 0.7500 1.0000 0.7500 0.5625
[4,] 0.4219 0.5625 0.7500 1.0000 0.7500
[5,] 0.3164 0.4219 0.5625 0.7500 1.0000
dp1$alpha :  6  3  0 -6 -3 
mix.p : 1 
\end{verbatim}
\begin{table}[ht]
\centering
\caption{Error quantiles for a fixed grid of points} 
\begin{tabular}{lllllll}
  \hline
 & 25\% & 50\% & 75\% & 90\% & 95\% & 99\% \\ 
  \hline
kde & 3.71E-09 & 8.1E-05 & 1.58E+10 & 4.49E+58 & 1.18E+89 &  Inf \\ 
  wkde & 3.58E-09 & 7.9E-05 & 1.51E+10 & 4.04E+58 & 1.14E+89 &  Inf \\ 
  wkdeA & 3.65E-09 & 8.01E-05 & 1.53E+10 & 4.28E+58 & 1.14E+89 &  Inf \\ 
  fill+wkde & 5.88E-07 & 0.00162 & 1.1E+13 & 5.75E+64 & 1.33E+98 &  Inf \\ 
   \hline
\end{tabular}
\end{table}
\begin{table}[ht]
\centering
\caption{Error quantiles evaluating at the observed sample points} 
\begin{tabular}{lllllll}
  \hline
 & 25\% & 50\% & 75\% & 90\% & 95\% & 99\% \\ 
  \hline
kde & 0.0361 & 0.0783 & 0.149 & 0.284 & 0.435 & 1.09 \\ 
  wkde & 0.0355 & 0.0773 & 0.148 & 0.285 & 0.438 & 1.11 \\ 
  wkdeA & 0.036 & 0.0779 & 0.149 & 0.285 & 0.437 & 1.11 \\ 
  fill+wkde & 0.0345 & 0.0748 & 0.141 & 0.268 & 0.414 & 1.05 \\ 
   \hline
\end{tabular}
\end{table}
\clearpage\centerline{\Large Case No.\, 35 }
\begin{verbatim}
Niter : 2500 
n : 250 
d : 4 
m : 3 
qN : 3 
Npts : 2401 
dp1$xi : 0 0 0 0 
dp1$Omega : 
     [,1] [,2] [,3] [,4]
[1,]    1    0    0    0
[2,]    0    1    0    0
[3,]    0    0    1    0
[4,]    0    0    0    1
dp1$alpha : 0 0 0 0 
mix.p : 1 
\end{verbatim}
\begin{table}[ht]
\centering
\caption{Error quantiles for a fixed grid of points} 
\begin{tabular}{lllllll}
  \hline
 & 25\% & 50\% & 75\% & 90\% & 95\% & 99\% \\ 
  \hline
kde & 0.000683 & 0.00247 & 0.00828 & 0.0239 & 0.0424 & 0.119 \\ 
  wkde & 0.000686 & 0.00248 & 0.00826 & 0.0236 & 0.0417 & 0.119 \\ 
  wkdeA & 0.000684 & 0.00247 & 0.00828 & 0.0238 & 0.0422 & 0.119 \\ 
  fill+wkde & 0.000698 & 0.00282 & 0.0098 & 0.0288 & 0.0502 & 0.134 \\ 
   \hline
\end{tabular}
\end{table}
\begin{table}[ht]
\centering
\caption{Error quantiles evaluating at the observed sample points} 
\begin{tabular}{lllllll}
  \hline
 & 25\% & 50\% & 75\% & 90\% & 95\% & 99\% \\ 
  \hline
kde & 0.0222 & 0.049 & 0.0943 & 0.165 & 0.243 & 0.579 \\ 
  wkde & 0.022 & 0.049 & 0.0951 & 0.167 & 0.245 & 0.582 \\ 
  wkdeA & 0.0222 & 0.0494 & 0.0949 & 0.166 & 0.243 & 0.578 \\ 
  fill+wkde & 0.0184 & 0.041 & 0.0825 & 0.15 & 0.221 & 0.522 \\ 
   \hline
\end{tabular}
\end{table}
\clearpage\centerline{\Large Case No.\, 36 }
\begin{verbatim}
Niter : 2500 
n : 250 
d : 4 
m : 2 
qN : 3 
Npts : 2401 
dp1$xi : 0 0 0 0 
dp1$Omega : 
     [,1] [,2] [,3] [,4]
[1,]    1    0    0    0
[2,]    0    1    0    0
[3,]    0    0    1    0
[4,]    0    0    0    1
dp1$alpha : 0 0 0 0 
mix.p : 1 
\end{verbatim}
\begin{table}[ht]
\centering
\caption{Error quantiles for a fixed grid of points} 
\begin{tabular}{lllllll}
  \hline
 & 25\% & 50\% & 75\% & 90\% & 95\% & 99\% \\ 
  \hline
kde & 0.000685 & 0.00247 & 0.00826 & 0.0238 & 0.0423 & 0.119 \\ 
  wkde & 0.000691 & 0.00247 & 0.00807 & 0.0224 & 0.0388 & 0.11 \\ 
  wkdeA & 0.000687 & 0.00245 & 0.00808 & 0.0225 & 0.0391 & 0.109 \\ 
  fill+wkde & 0.000857 & 0.00372 & 0.014 & 0.0398 & 0.0674 & 0.176 \\ 
   \hline
\end{tabular}
\end{table}
\begin{table}[ht]
\centering
\caption{Error quantiles evaluating at the observed sample points} 
\begin{tabular}{lllllll}
  \hline
 & 25\% & 50\% & 75\% & 90\% & 95\% & 99\% \\ 
  \hline
kde & 0.0225 & 0.0493 & 0.0941 & 0.165 & 0.243 & 0.586 \\ 
  wkde & 0.0216 & 0.0478 & 0.0909 & 0.157 & 0.229 & 0.542 \\ 
  wkdeA & 0.0226 & 0.0486 & 0.0899 & 0.154 & 0.225 & 0.535 \\ 
  fill+wkde & 0.0145 & 0.0312 & 0.0595 & 0.112 & 0.168 & 0.39 \\ 
   \hline
\end{tabular}
\end{table}
\clearpage\centerline{\Large Case No.\, 37 }
\begin{verbatim}
Niter : 2500 
n : 1000 
d : 4 
m : 2 
qN : 3 
Npts : 2401 
dp1$xi : 0 0 0 0 
dp1$Omega : 
     [,1] [,2] [,3] [,4]
[1,]    1    0    0    0
[2,]    0    1    0    0
[3,]    0    0    1    0
[4,]    0    0    0    1
dp1$alpha : 0 0 0 0 
mix.p : 1 
\end{verbatim}
\begin{table}[ht]
\centering
\caption{Error quantiles for a fixed grid of points} 
\begin{tabular}{lllllll}
  \hline
 & 25\% & 50\% & 75\% & 90\% & 95\% & 99\% \\ 
  \hline
kde & 0.000652 & 0.00225 & 0.00714 & 0.019 & 0.0321 & 0.0787 \\ 
  wkde & 0.000654 & 0.00224 & 0.00676 & 0.0164 & 0.0268 & 0.0644 \\ 
  wkdeA & 0.000653 & 0.00224 & 0.00673 & 0.0164 & 0.0268 & 0.0637 \\ 
  fill+wkde & 0.00225 & 0.00766 & 0.0214 & 0.0467 & 0.0718 & 0.149 \\ 
   \hline
\end{tabular}
\end{table}
\begin{table}[ht]
\centering
\caption{Error quantiles evaluating at the observed sample points} 
\begin{tabular}{lllllll}
  \hline
 & 25\% & 50\% & 75\% & 90\% & 95\% & 99\% \\ 
  \hline
kde & 0.0111 & 0.024 & 0.045 & 0.0779 & 0.112 & 0.262 \\ 
  wkde & 0.0108 & 0.0235 & 0.0433 & 0.0699 & 0.095 & 0.209 \\ 
  wkdeA & 0.0107 & 0.023 & 0.0414 & 0.0667 & 0.0922 & 0.208 \\ 
  fill+wkde & 0.00844 & 0.0177 & 0.0303 & 0.0491 & 0.0729 & 0.156 \\ 
   \hline
\end{tabular}
\end{table}
\clearpage\centerline{\Large Case No.\, 38 }
\begin{verbatim}
Niter : 2500 
n : 250 
d : 4 
m : 2 
qN : 6 
Npts : 1296 
dp1$xi : 2 2 2 2 
dp1$Omega : 
         [,1]     [,2]     [,3]     [,4]
[1,]  1.00000  0.60714 -0.04464 -0.33705
[2,]  0.60714  1.00000  0.60714 -0.04464
[3,] -0.04464  0.60714  1.00000  0.60714
[4,] -0.33705 -0.04464  0.60714  1.00000
dp1$alpha :  6  3 -6 -3 
dp1$nu : 5 
mix.p : 0.3333 
dp2$xi : -2 -2 -2 -2 
dp2$Omega : 
       [,1]   [,2]   [,3]   [,4]
[1,] 1.0000 0.7500 0.5625 0.4219
[2,] 0.7500 1.0000 0.7500 0.5625
[3,] 0.5625 0.7500 1.0000 0.7500
[4,] 0.4219 0.5625 0.7500 1.0000
dp2$alpha : -6 -3  6  3 
dp2$nu : 5 
\end{verbatim}
\begin{table}[ht]
\centering
\caption{Error quantiles for a fixed grid of points} 
\begin{tabular}{lllllll}
  \hline
 & 25\% & 50\% & 75\% & 90\% & 95\% & 99\% \\ 
  \hline
kde & 2.89E-05 & 9.71E-05 & 0.000349 & 0.00119 & 0.00318 & 0.0473 \\ 
  wkde & 2.89E-05 & 9.71E-05 & 0.000349 & 0.00118 & 0.00307 & 0.0445 \\ 
  wkdeA & 2.89E-05 & 9.71E-05 & 0.000349 & 0.00118 & 0.00309 & 0.0446 \\ 
  fill+wkde & 2.9E-05 & 9.93E-05 & 0.000372 & 0.00168 & 0.00613 & 0.0651 \\ 
   \hline
\end{tabular}
\end{table}
\begin{table}[ht]
\centering
\caption{Error quantiles evaluating at the observed sample points} 
\begin{tabular}{lllllll}
  \hline
 & 25\% & 50\% & 75\% & 90\% & 95\% & 99\% \\ 
  \hline
kde & 0.0309 & 0.0677 & 0.127 & 0.199 & 0.25 & 0.708 \\ 
  wkde & 0.0297 & 0.0654 & 0.124 & 0.194 & 0.243 & 0.595 \\ 
  wkdeA & 0.0301 & 0.0658 & 0.124 & 0.194 & 0.243 & 0.605 \\ 
  fill+wkde & 0.0295 & 0.0652 & 0.124 & 0.194 & 0.242 & 0.569 \\ 
   \hline
\end{tabular}
\end{table}
\clearpage\centerline{\Large Case No.\, 39 }
\begin{verbatim}
Niter : 2500 
n : 250 
d : 4 
m : 2 
qN : 6 
Npts : 1296 
dp1$xi : 2 2 2 2 
dp1$Omega : 
         [,1]     [,2]     [,3]     [,4]
[1,]  1.00000  0.60714 -0.04464 -0.33705
[2,]  0.60714  1.00000  0.60714 -0.04464
[3,] -0.04464  0.60714  1.00000  0.60714
[4,] -0.33705 -0.04464  0.60714  1.00000
dp1$alpha :  6  3 -6 -3 
dp1$nu : 2 
mix.p : 0.3333 
dp2$xi : -2 -2 -2 -2 
dp2$Omega : 
       [,1]   [,2]   [,3]   [,4]
[1,] 1.0000 0.7500 0.5625 0.4219
[2,] 0.7500 1.0000 0.7500 0.5625
[3,] 0.5625 0.7500 1.0000 0.7500
[4,] 0.4219 0.5625 0.7500 1.0000
dp2$alpha : -6 -3  6  3 
dp2$nu : 2 
\end{verbatim}
\begin{table}[ht]
\centering
\caption{Error quantiles for a fixed grid of points} 
\begin{tabular}{lllllll}
  \hline
 & 25\% & 50\% & 75\% & 90\% & 95\% & 99\% \\ 
  \hline
kde & 0.000209 & 0.00049 & 0.00127 & 0.00378 & 0.0102 & 0.0589 \\ 
  wkde & 0.000209 & 0.000491 & 0.00126 & 0.00352 & 0.00854 & 0.0527 \\ 
  wkdeA & 0.000209 & 0.00049 & 0.00126 & 0.00354 & 0.00866 & 0.0526 \\ 
  fill+wkde & 0.000205 & 0.00049 & 0.0013 & 0.00416 & 0.0115 & 0.0606 \\ 
   \hline
\end{tabular}
\end{table}
\begin{table}[ht]
\centering
\caption{Error quantiles evaluating at the observed sample points} 
\begin{tabular}{lllllll}
  \hline
 & 25\% & 50\% & 75\% & 90\% & 95\% & 99\% \\ 
  \hline
kde & 0.0282 & 0.0663 & 0.148 & 0.259 & 0.366 & 2.95 \\ 
  wkde & 0.0233 & 0.0589 & 0.139 & 0.248 & 0.339 & 2.99 \\ 
  wkdeA & 0.0242 & 0.0593 & 0.139 & 0.249 & 0.341 &    3 \\ 
  fill+wkde & 0.0225 & 0.0577 & 0.138 & 0.244 & 0.323 & 2.16 \\ 
   \hline
\end{tabular}
\end{table}
\clearpage\centerline{\Large Case No.\, 40 }
\begin{verbatim}
Niter : 2500 
n : 250 
d : 4 
m : 2 
qN : 3 
Npts : 2401 
dp1$xi : 0 0 0 0 
dp1$Omega : 
     [,1] [,2] [,3] [,4]
[1,]    1    0    0    0
[2,]    0    1    0    0
[3,]    0    0    1    0
[4,]    0    0    0    1
dp1$alpha :  6  3 -6 -3 
mix.p : 1 
\end{verbatim}
\begin{table}[ht]
\centering
\caption{Error quantiles for a fixed grid of points} 
\begin{tabular}{lllllll}
  \hline
 & 25\% & 50\% & 75\% & 90\% & 95\% & 99\% \\ 
  \hline
kde & 0.00198 & 0.016 & 4.96E+11 & 3.54E+51 & 3.73E+82 &  Inf \\ 
  wkde & 0.00198 & 0.0156 & 4.06E+11 & 2.84E+51 & 2.83E+82 &  Inf \\ 
  wkdeA & 0.00198 & 0.0156 & 4.34E+11 & 3.07E+51 & 3.15E+82 &  Inf \\ 
  fill+wkde & 0.00243 & 0.0278 & 9.96E+13 & 4.85E+58 & 4.97E+93 &  Inf \\ 
   \hline
\end{tabular}
\end{table}
\begin{table}[ht]
\centering
\caption{Error quantiles evaluating at the observed sample points} 
\begin{tabular}{lllllll}
  \hline
 & 25\% & 50\% & 75\% & 90\% & 95\% & 99\% \\ 
  \hline
kde & 0.0264 & 0.0573 & 0.11 & 0.197 & 0.292 & 0.696 \\ 
  wkde & 0.0251 & 0.0554 & 0.107 & 0.193 & 0.284 & 0.679 \\ 
  wkdeA & 0.0257 & 0.0559 & 0.106 & 0.189 & 0.278 & 0.663 \\ 
  fill+wkde & 0.0201 & 0.0435 & 0.0803 & 0.142 & 0.214 & 0.509 \\ 
   \hline
\end{tabular}
\end{table}
\clearpage\centerline{\Large Case No.\, 41 }
\begin{verbatim}
Niter : 2500 
n : 1000 
d : 4 
m : 2 
qN : 3 
Npts : 2401 
dp1$xi : 0 0 0 0 
dp1$Omega : 
     [,1] [,2] [,3] [,4]
[1,]    1    0    0    0
[2,]    0    1    0    0
[3,]    0    0    1    0
[4,]    0    0    0    1
dp1$alpha :  6  3 -6 -3 
mix.p : 1 
\end{verbatim}
\begin{table}[ht]
\centering
\caption{Error quantiles for a fixed grid of points} 
\begin{tabular}{lllllll}
  \hline
 & 25\% & 50\% & 75\% & 90\% & 95\% & 99\% \\ 
  \hline
kde & 0.00189 & 0.0135 & 3.41E+10 & 2.95E+47 & 1.92E+76 &  Inf \\ 
  wkde & 0.0019 & 0.013 & 2.43E+10 & 2.13E+47 & 1.38E+76 &  Inf \\ 
  wkdeA & 0.00189 & 0.013 & 2.58E+10 & 2.27E+47 & 1.45E+76 &  Inf \\ 
  fill+wkde & 0.00348 & 0.0329 & 1.26E+14 & 5.02E+58 & 4.16E+93 &  Inf \\ 
   \hline
\end{tabular}
\end{table}
\begin{table}[ht]
\centering
\caption{Error quantiles evaluating at the observed sample points} 
\begin{tabular}{lllllll}
  \hline
 & 25\% & 50\% & 75\% & 90\% & 95\% & 99\% \\ 
  \hline
kde & 0.0154 & 0.0332 & 0.0604 & 0.102 & 0.148 & 0.351 \\ 
  wkde & 0.015 & 0.0327 & 0.0599 & 0.0983 & 0.138 & 0.316 \\ 
  wkdeA & 0.015 & 0.0324 & 0.0581 & 0.0946 & 0.134 & 0.311 \\ 
  fill+wkde & 0.0128 & 0.028 & 0.0517 & 0.0801 & 0.105 & 0.235 \\ 
   \hline
\end{tabular}
\end{table}
\clearpage\centerline{\Large Case No.\, 42 }
\begin{verbatim}
Niter : 2500 
n : 250 
d : 4 
m : 2 
qN : 3 
Npts : 2401 
dp1$xi : 0 0 0 0 
dp1$Omega : 
       [,1]   [,2]   [,3]   [,4]
[1,] 1.0000 0.7500 0.5625 0.4219
[2,] 0.7500 1.0000 0.7500 0.5625
[3,] 0.5625 0.7500 1.0000 0.7500
[4,] 0.4219 0.5625 0.7500 1.0000
dp1$alpha :  6  3 -6 -3 
mix.p : 1 
\end{verbatim}
\begin{table}[ht]
\centering
\caption{Error quantiles for a fixed grid of points} 
\begin{tabular}{lllllll}
  \hline
 & 25\% & 50\% & 75\% & 90\% & 95\% & 99\% \\ 
  \hline
kde & 1.51E-06 & 0.00317 & 6.56E+09 & 2.19E+49 & 9.61E+79 &  Inf \\ 
  wkde & 1.3E-06 & 0.00267 & 4.83E+09 & 1.74E+49 & 7.73E+79 &  Inf \\ 
  wkdeA & 1.34E-06 & 0.00279 & 5.23E+09 & 1.84E+49 & 7.78E+79 &  Inf \\ 
  fill+wkde & 0.0011 & 0.096 & 3.54E+13 & 5.81E+57 & 1.99E+93 &  Inf \\ 
   \hline
\end{tabular}
\end{table}
\begin{table}[ht]
\centering
\caption{Error quantiles evaluating at the observed sample points} 
\begin{tabular}{lllllll}
  \hline
 & 25\% & 50\% & 75\% & 90\% & 95\% & 99\% \\ 
  \hline
kde & 0.0366 & 0.078 & 0.138 & 0.224 & 0.323 & 0.739 \\ 
  wkde & 0.0334 & 0.0718 & 0.129 & 0.21 & 0.302 & 0.693 \\ 
  wkdeA & 0.0335 & 0.0717 & 0.128 & 0.207 & 0.297 & 0.68 \\ 
  fill+wkde & 0.033 & 0.0707 & 0.125 & 0.194 & 0.271 & 0.627 \\ 
   \hline
\end{tabular}
\end{table}
\clearpage\centerline{\Large Case No.\, 43 }
\begin{verbatim}
Niter : 2500 
n : 1000 
d : 4 
m : 2 
qN : 3 
Npts : 2401 
dp1$xi : 0 0 0 0 
dp1$Omega : 
       [,1]   [,2]   [,3]   [,4]
[1,] 1.0000 0.7500 0.5625 0.4219
[2,] 0.7500 1.0000 0.7500 0.5625
[3,] 0.5625 0.7500 1.0000 0.7500
[4,] 0.4219 0.5625 0.7500 1.0000
dp1$alpha :  6  3 -6 -3 
mix.p : 1 
\end{verbatim}
\begin{table}[ht]
\centering
\caption{Error quantiles for a fixed grid of points} 
\begin{tabular}{lllllll}
  \hline
 & 25\% & 50\% & 75\% & 90\% & 95\% & 99\% \\ 
  \hline
kde & 2.78E-07 & 0.000777 & 5.58E+04 & 1.12E+40 & 1.22E+67 &  Inf \\ 
  wkde & 2.53E-07 & 0.000672 & 3E+04 & 6.66E+39 & 1.09E+67 &  Inf \\ 
  wkdeA & 2.57E-07 & 0.000686 & 3.3E+04 & 7.05E+39 & 1.12E+67 &  Inf \\ 
  fill+wkde & 0.000119 & 0.027 & 1.64E+11 & 4.85E+53 & 1.44E+87 &  Inf \\ 
   \hline
\end{tabular}
\end{table}
\begin{table}[ht]
\centering
\caption{Error quantiles evaluating at the observed sample points} 
\begin{tabular}{lllllll}
  \hline
 & 25\% & 50\% & 75\% & 90\% & 95\% & 99\% \\ 
  \hline
kde & 0.0264 & 0.0563 & 0.0991 & 0.153 & 0.207 & 0.462 \\ 
  wkde & 0.0225 & 0.0487 & 0.0877 & 0.137 & 0.185 & 0.417 \\ 
  wkdeA & 0.0222 & 0.048 & 0.0858 & 0.134 & 0.181 & 0.41 \\ 
  fill+wkde & 0.0228 & 0.0494 & 0.0887 & 0.135 & 0.172 & 0.373 \\ 
   \hline
\end{tabular}
\end{table}
\clearpage\centerline{\Large Case No.\, 44 }
\begin{verbatim}
Niter : 2500 
n : 1000 
d : 5 
m : 2 
qN : 3 
Npts : 3125 
dp1$xi : 0 0 0 0 0 
dp1$Omega : 
       [,1]   [,2]   [,3]   [,4]   [,5]
[1,] 1.0000 0.7500 0.5625 0.4219 0.3164
[2,] 0.7500 1.0000 0.7500 0.5625 0.4219
[3,] 0.5625 0.7500 1.0000 0.7500 0.5625
[4,] 0.4219 0.5625 0.7500 1.0000 0.7500
[5,] 0.3164 0.4219 0.5625 0.7500 1.0000
dp1$alpha :  6  3  0 -6 -3 
mix.p : 1 
\end{verbatim}
\begin{table}[ht]
\centering
\caption{Error quantiles for a fixed grid of points} 
\begin{tabular}{lllllll}
  \hline
 & 25\% & 50\% & 75\% & 90\% & 95\% & 99\% \\ 
  \hline
kde & 1.79E-10 & 5.72E-06 & 1.17E+04 & 4.81E+46 & 3.23E+78 &  Inf \\ 
  wkde & 1.46E-10 & 5.02E-06 & 4.71E+03 & 1.16E+46 & 1.74E+78 &  Inf \\ 
  wkdeA & 1.53E-10 & 5.17E-06 & 5.95E+03 & 1.59E+46 & 1.96E+78 &  Inf \\ 
  fill+wkde & 0.000183 & 0.0602 & 3.59E+14 & 3.77E+67 & 6.75E+99 &  Inf \\ 
   \hline
\end{tabular}
\end{table}
\begin{table}[ht]
\centering
\caption{Error quantiles evaluating at the observed sample points} 
\begin{tabular}{lllllll}
  \hline
 & 25\% & 50\% & 75\% & 90\% & 95\% & 99\% \\ 
  \hline
kde & 0.0267 & 0.0577 & 0.106 & 0.188 & 0.287 & 0.73 \\ 
  wkde & 0.0203 & 0.0455 & 0.0883 & 0.163 & 0.254 & 0.667 \\ 
  wkdeA & 0.0208 & 0.0459 & 0.0875 & 0.161 & 0.252 & 0.66 \\ 
  fill+wkde & 0.0196 & 0.0433 & 0.08 & 0.13 & 0.187 & 0.501 \\ 
   \hline
\end{tabular}
\end{table}
\clearpage\centerline{\Large Case No.\, 45 }
\begin{verbatim}
Niter : 5000 
n : 250 
d : 5 
m : 2 
qN : 3 
Npts : 3125 
dp1$xi : 0 0 0 0 0 
dp1$Omega : 
       [,1]   [,2]   [,3]   [,4]   [,5]
[1,] 1.0000 0.7500 0.5625 0.4219 0.3164
[2,] 0.7500 1.0000 0.7500 0.5625 0.4219
[3,] 0.5625 0.7500 1.0000 0.7500 0.5625
[4,] 0.4219 0.5625 0.7500 1.0000 0.7500
[5,] 0.3164 0.4219 0.5625 0.7500 1.0000
dp1$alpha :  6  3  0 -6 -3 
mix.p : 1 
\end{verbatim}
\begin{table}[ht]
\centering
\caption{Error quantiles for a fixed grid of points} 
\begin{tabular}{lllllll}
  \hline
 & 25\% & 50\% & 75\% & 90\% & 95\% & 99\% \\ 
  \hline
kde & 3.71E-09 & 8.17E-05 & 1.11E+10 & 2.47E+58 & 1.55E+89 &  Inf \\ 
  wkde & 2.76E-09 & 6.56E-05 & 6.99E+09 & 1.34E+58 & 1.15E+89 &  Inf \\ 
  wkdeA & 3.07E-09 & 7.12E-05 & 8.5E+09 & 1.7E+58 & 1.22E+89 &  Inf \\ 
  fill+wkde & 0.0014 & 0.695 & 1.19E+17 & 6.06E+71 & 1.77E+105 &  Inf \\ 
   \hline
\end{tabular}
\end{table}
\begin{table}[ht]
\centering
\caption{Error quantiles evaluating at the observed sample points} 
\begin{tabular}{lllllll}
  \hline
 & 25\% & 50\% & 75\% & 90\% & 95\% & 99\% \\ 
  \hline
kde & 0.0361 & 0.0784 & 0.15 & 0.284 & 0.436 & 1.11 \\ 
  wkde & 0.0314 & 0.0692 & 0.135 & 0.263 & 0.41 & 1.06 \\ 
  wkdeA & 0.0326 & 0.0713 & 0.137 & 0.263 & 0.408 & 1.05 \\ 
  fill+wkde & 0.0291 & 0.0632 & 0.116 & 0.206 & 0.324 & 0.838 \\ 
   \hline
\end{tabular}
\end{table}
\clearpage\centerline{\Large Case No.\, 46 }
\begin{verbatim}
Niter : 10000 
n : 250 
d : 4 
m : 2 
qN : 6 
Npts : 2401 
dp1$xi : 2 2 2 2 
dp1$Omega : 
         [,1]     [,2]     [,3]     [,4]
[1,]  1.00000  0.60714 -0.04464 -0.33705
[2,]  0.60714  1.00000  0.60714 -0.04464
[3,] -0.04464  0.60714  1.00000  0.60714
[4,] -0.33705 -0.04464  0.60714  1.00000
dp1$alpha :  6  3 -6 -3 
mix.p : 0.3333 
dp2$xi : -2 -2 -2 -2 
dp2$Omega : 
       [,1]   [,2]   [,3]   [,4]
[1,] 1.0000 0.7500 0.5625 0.4219
[2,] 0.7500 1.0000 0.7500 0.5625
[3,] 0.5625 0.7500 1.0000 0.7500
[4,] 0.4219 0.5625 0.7500 1.0000
dp2$alpha : -6 -3  6  3 
\end{verbatim}
\begin{table}[ht]
\centering
\caption{Error quantiles for a fixed grid of points} 
\begin{tabular}{lllllll}
  \hline
 & 25\% & 50\% & 75\% & 90\% & 95\% & 99\% \\ 
  \hline
kde & 5.63E-14 & 4.35E-07 & 0.112 & 3.97E+11 & 1.62E+25 & 1.01E+55 \\ 
  wkde & 5.57E-14 & 4.32E-07 & 0.11 & 3.82E+11 & 1.33E+25 & 1.01E+55 \\ 
  wkdeA & 5.55E-14 & 4.29E-07 & 0.11 & 3.82E+11 & 1.51E+25 & 9.41E+54 \\ 
  fill+wkde & 6.93E-11 & 0.000103 & 3.38E+03 & 1.09E+19 & 2.37E+37 & 2.28E+73 \\ 
   \hline
\end{tabular}
\end{table}
\begin{table}[ht]
\centering
\caption{Error quantiles evaluating at the observed sample points} 
\begin{tabular}{lllllll}
  \hline
 & 25\% & 50\% & 75\% & 90\% & 95\% & 99\% \\ 
  \hline
kde & 0.0307 & 0.0652 & 0.112 & 0.157 & 0.185 & 0.26 \\ 
  wkde & 0.0307 & 0.0651 & 0.112 & 0.157 & 0.185 & 0.261 \\ 
  wkdeA & 0.0305 & 0.0648 & 0.111 & 0.156 & 0.184 & 0.258 \\ 
  fill+wkde & 0.0308 & 0.0654 & 0.112 & 0.158 & 0.185 & 0.26 \\ 
   \hline
\end{tabular}
\end{table}
\clearpage\centerline{\Large Case No.\, 47 }
\begin{verbatim}
Niter : 2500 
n : 1000 
d : 4 
m : 2 
qN : 6 
Npts : 2401 
dp1$xi : 2 2 2 2 
dp1$Omega : 
         [,1]     [,2]     [,3]     [,4]
[1,]  1.00000  0.60714 -0.04464 -0.33705
[2,]  0.60714  1.00000  0.60714 -0.04464
[3,] -0.04464  0.60714  1.00000  0.60714
[4,] -0.33705 -0.04464  0.60714  1.00000
dp1$alpha :  6  3 -6 -3 
mix.p : 0.3333 
dp2$xi : -2 -2 -2 -2 
dp2$Omega : 
       [,1]   [,2]   [,3]   [,4]
[1,] 1.0000 0.7500 0.5625 0.4219
[2,] 0.7500 1.0000 0.7500 0.5625
[3,] 0.5625 0.7500 1.0000 0.7500
[4,] 0.4219 0.5625 0.7500 1.0000
dp2$alpha : -6 -3  6  3 
\end{verbatim}
\begin{table}[ht]
\centering
\caption{Error quantiles for a fixed grid of points} 
\begin{tabular}{lllllll}
  \hline
 & 25\% & 50\% & 75\% & 90\% & 95\% & 99\% \\ 
  \hline
kde & 1.73E-25 & 6.07E-15 & 1.58E-07 & 0.0214 &  160 & 2.42E+15 \\ 
  wkde & 1.66E-25 & 5.72E-15 & 1.5E-07 & 0.0201 &  145 & 8.08E+14 \\ 
  wkdeA & 1.68E-25 & 5.84E-15 & 1.53E-07 & 0.0206 &  150 & 1.44E+15 \\ 
  fill+wkde & 7.64E-20 & 7.15E-11 & 0.00147 & 2.77E+08 & 5.19E+21 & 1.24E+49 \\ 
   \hline
\end{tabular}
\end{table}
\begin{table}[ht]
\centering
\caption{Error quantiles evaluating at the observed sample points} 
\begin{tabular}{lllllll}
  \hline
 & 25\% & 50\% & 75\% & 90\% & 95\% & 99\% \\ 
  \hline
kde & 0.0229 & 0.0492 & 0.0857 & 0.124 & 0.148 & 0.209 \\ 
  wkde & 0.0225 & 0.0483 & 0.0844 & 0.122 & 0.146 & 0.207 \\ 
  wkdeA & 0.0226 & 0.0485 & 0.0846 & 0.122 & 0.146 & 0.205 \\ 
  fill+wkde & 0.0226 & 0.0486 & 0.085 & 0.123 & 0.147 & 0.206 \\ 
   \hline
\end{tabular}
\end{table}
\clearpage\centerline{\Large Case No.\, 48 }
\begin{verbatim}
Niter : 2500 
n : 250 
d : 4 
m : 2 
qN : 3 
Npts : 2401 
dp1$xi : 0 0 0 0 
dp1$Omega : 
       [,1]   [,2]   [,3]   [,4]
[1,] 1.0000 0.7500 0.5625 0.4219
[2,] 0.7500 1.0000 0.7500 0.5625
[3,] 0.5625 0.7500 1.0000 0.7500
[4,] 0.4219 0.5625 0.7500 1.0000
dp1$alpha : 0 0 0 0 
dp1$nu : 5 
mix.p : 1 
\end{verbatim}
\begin{table}[ht]
\centering
\caption{Error quantiles for a fixed grid of points} 
\begin{tabular}{lllllll}
  \hline
 & 25\% & 50\% & 75\% & 90\% & 95\% & 99\% \\ 
  \hline
kde & 0.000881 & 0.00218 & 0.00624 & 0.02 & 0.0424 & 0.123 \\ 
  wkde & 0.000897 & 0.00221 & 0.00587 & 0.016 & 0.0319 &  0.1 \\ 
  wkdeA & 0.000893 & 0.0022 & 0.00586 & 0.0163 & 0.0325 & 0.0983 \\ 
  fill+wkde & 0.000826 & 0.00265 & 0.0106 & 0.037 & 0.0715 & 0.199 \\ 
   \hline
\end{tabular}
\end{table}
\begin{table}[ht]
\centering
\caption{Error quantiles evaluating at the observed sample points} 
\begin{tabular}{lllllll}
  \hline
 & 25\% & 50\% & 75\% & 90\% & 95\% & 99\% \\ 
  \hline
kde & 0.0358 & 0.0763 & 0.137 & 0.25 & 0.461 & 2.11 \\ 
  wkde & 0.0286 & 0.0629 & 0.118 & 0.221 & 0.42 & 2.09 \\ 
  wkdeA & 0.0303 & 0.0647 & 0.118 & 0.22 & 0.421 & 2.09 \\ 
  fill+wkde & 0.0259 & 0.0575 & 0.109 & 0.181 & 0.295 & 1.37 \\ 
   \hline
\end{tabular}
\end{table}
\clearpage\centerline{\Large Case No.\, 49 }
\begin{verbatim}
Niter : 2500 
n : 500 
d : 4 
m : 2 
qN : 3 
Npts : 2401 
dp1$xi : 0 0 0 0 
dp1$Omega : 
       [,1]   [,2]   [,3]   [,4]
[1,] 1.0000 0.7500 0.5625 0.4219
[2,] 0.7500 1.0000 0.7500 0.5625
[3,] 0.5625 0.7500 1.0000 0.7500
[4,] 0.4219 0.5625 0.7500 1.0000
dp1$alpha : 0 0 0 0 
dp1$nu : 5 
mix.p : 1 
\end{verbatim}
\begin{table}[ht]
\centering
\caption{Error quantiles for a fixed grid of points} 
\begin{tabular}{lllllll}
  \hline
 & 25\% & 50\% & 75\% & 90\% & 95\% & 99\% \\ 
  \hline
kde & 0.00089 & 0.00221 & 0.00612 & 0.018 & 0.0374 & 0.108 \\ 
  wkde & 0.000908 & 0.00227 & 0.00588 & 0.0141 & 0.0257 & 0.0839 \\ 
  wkdeA & 0.000906 & 0.00226 & 0.00583 & 0.0141 & 0.0257 & 0.0815 \\ 
  fill+wkde & 0.000833 & 0.00229 & 0.00728 & 0.0231 & 0.0462 & 0.135 \\ 
   \hline
\end{tabular}
\end{table}
\begin{table}[ht]
\centering
\caption{Error quantiles evaluating at the observed sample points} 
\begin{tabular}{lllllll}
  \hline
 & 25\% & 50\% & 75\% & 90\% & 95\% & 99\% \\ 
  \hline
kde & 0.0305 & 0.0651 & 0.116 & 0.204 & 0.375 & 1.62 \\ 
  wkde & 0.0217 & 0.0495 & 0.0956 & 0.178 & 0.342 & 1.73 \\ 
  wkdeA & 0.0226 & 0.0501 & 0.0947 & 0.175 & 0.34 & 1.72 \\ 
  fill+wkde & 0.02 & 0.0452 & 0.0878 & 0.148 & 0.215 & 0.992 \\ 
   \hline
\end{tabular}
\end{table}
\clearpage\centerline{\Large Case No.\, 50 }
\begin{verbatim}
Niter : 2500 
n : 1000 
d : 4 
m : 2 
qN : 3 
Npts : 2401 
dp1$xi : 0 0 0 0 
dp1$Omega : 
       [,1]   [,2]   [,3]   [,4]
[1,] 1.0000 0.7500 0.5625 0.4219
[2,] 0.7500 1.0000 0.7500 0.5625
[3,] 0.5625 0.7500 1.0000 0.7500
[4,] 0.4219 0.5625 0.7500 1.0000
dp1$alpha : 0 0 0 0 
dp1$nu : 5 
mix.p : 1 
\end{verbatim}
\begin{table}[ht]
\centering
\caption{Error quantiles for a fixed grid of points} 
\begin{tabular}{lllllll}
  \hline
 & 25\% & 50\% & 75\% & 90\% & 95\% & 99\% \\ 
  \hline
kde & 0.000894 & 0.00223 & 0.00603 & 0.0166 & 0.0333 & 0.0951 \\ 
  wkde & 0.000914 & 0.00232 & 0.00601 & 0.0134 & 0.0224 & 0.0737 \\ 
  wkdeA & 0.000912 & 0.0023 & 0.00594 & 0.0132 & 0.0222 & 0.0713 \\ 
  fill+wkde & 0.000863 & 0.00212 & 0.00569 & 0.014 & 0.0235 & 0.059 \\ 
   \hline
\end{tabular}
\end{table}
\begin{table}[ht]
\centering
\caption{Error quantiles evaluating at the observed sample points} 
\begin{tabular}{lllllll}
  \hline
 & 25\% & 50\% & 75\% & 90\% & 95\% & 99\% \\ 
  \hline
kde & 0.0259 & 0.0549 & 0.0978 & 0.167 &  0.3 & 1.33 \\ 
  wkde & 0.0166 & 0.0392 & 0.0786 & 0.148 & 0.285 & 1.52 \\ 
  wkdeA & 0.0168 & 0.0391 & 0.0773 & 0.145 & 0.282 &  1.5 \\ 
  fill+wkde & 0.0157 & 0.0361 & 0.0714 & 0.123 & 0.166 & 0.733 \\ 
   \hline
\end{tabular}
\end{table}
\clearpage\centerline{\Large Case No.\, 51 }
\begin{verbatim}
Niter : 2500 
n : 250 
d : 4 
m : 3 
qN : 3 
Npts : 2401 
dp1$xi : 0 0 0 0 
dp1$Omega : 
       [,1]   [,2]   [,3]   [,4]
[1,] 1.0000 0.7500 0.5625 0.4219
[2,] 0.7500 1.0000 0.7500 0.5625
[3,] 0.5625 0.7500 1.0000 0.7500
[4,] 0.4219 0.5625 0.7500 1.0000
dp1$alpha : 0 0 0 0 
dp1$nu : 5 
mix.p : 1 
\end{verbatim}
\begin{table}[ht]
\centering
\caption{Error quantiles for a fixed grid of points} 
\begin{tabular}{lllllll}
  \hline
 & 25\% & 50\% & 75\% & 90\% & 95\% & 99\% \\ 
  \hline
kde & 0.000882 & 0.00219 & 0.00623 & 0.0199 & 0.0421 & 0.123 \\ 
  wkde & 0.000882 & 0.00219 & 0.00623 & 0.0199 & 0.0422 & 0.124 \\ 
  wkdeA & 0.000882 & 0.00219 & 0.00623 & 0.0199 & 0.0421 & 0.124 \\ 
  fill+wkde & 0.000866 & 0.00215 & 0.00634 & 0.021 & 0.0436 & 0.125 \\ 
   \hline
\end{tabular}
\end{table}
\begin{table}[ht]
\centering
\caption{Error quantiles evaluating at the observed sample points} 
\begin{tabular}{lllllll}
  \hline
 & 25\% & 50\% & 75\% & 90\% & 95\% & 99\% \\ 
  \hline
kde & 0.0358 & 0.0763 & 0.137 & 0.251 & 0.46 & 2.08 \\ 
  wkde & 0.0357 & 0.0761 & 0.137 & 0.254 & 0.467 & 2.13 \\ 
  wkdeA & 0.0358 & 0.0762 & 0.137 & 0.253 & 0.465 & 2.12 \\ 
  fill+wkde & 0.0353 & 0.0753 & 0.136 & 0.248 & 0.455 & 2.07 \\ 
   \hline
\end{tabular}
\end{table}
\clearpage\centerline{\Large Case No.\, 52 }
\begin{verbatim}
Niter : 2500 
n : 500 
d : 4 
m : 3 
qN : 3 
Npts : 2401 
dp1$xi : 0 0 0 0 
dp1$Omega : 
       [,1]   [,2]   [,3]   [,4]
[1,] 1.0000 0.7500 0.5625 0.4219
[2,] 0.7500 1.0000 0.7500 0.5625
[3,] 0.5625 0.7500 1.0000 0.7500
[4,] 0.4219 0.5625 0.7500 1.0000
dp1$alpha : 0 0 0 0 
dp1$nu : 5 
mix.p : 1 
\end{verbatim}
\begin{table}[ht]
\centering
\caption{Error quantiles for a fixed grid of points} 
\begin{tabular}{lllllll}
  \hline
 & 25\% & 50\% & 75\% & 90\% & 95\% & 99\% \\ 
  \hline
kde & 0.00089 & 0.00221 & 0.00612 & 0.0181 & 0.0374 & 0.108 \\ 
  wkde & 0.00089 & 0.00221 & 0.00612 & 0.0181 & 0.0378 & 0.111 \\ 
  wkdeA & 0.00089 & 0.00221 & 0.00612 & 0.0181 & 0.0377 & 0.11 \\ 
  fill+wkde & 0.000883 & 0.00219 & 0.00613 & 0.0186 & 0.0382 & 0.11 \\ 
   \hline
\end{tabular}
\end{table}
\begin{table}[ht]
\centering
\caption{Error quantiles evaluating at the observed sample points} 
\begin{tabular}{lllllll}
  \hline
 & 25\% & 50\% & 75\% & 90\% & 95\% & 99\% \\ 
  \hline
kde & 0.0307 & 0.0652 & 0.116 & 0.204 & 0.372 & 1.64 \\ 
  wkde & 0.0305 & 0.0652 & 0.117 & 0.209 & 0.384 & 1.71 \\ 
  wkdeA & 0.0306 & 0.0651 & 0.117 & 0.208 & 0.382 &  1.7 \\ 
  fill+wkde & 0.03 & 0.0641 & 0.115 & 0.201 & 0.367 & 1.63 \\ 
   \hline
\end{tabular}
\end{table}
\clearpage\centerline{\Large Case No.\, 53 }
\begin{verbatim}
Niter : 2500 
n : 250 
d : 4 
m : 3 
qN : 3 
Npts : 2401 
dp1$xi : 0 0 0 0 
dp1$Omega : 
     [,1] [,2] [,3] [,4]
[1,]    1    0    0    0
[2,]    0    1    0    0
[3,]    0    0    1    0
[4,]    0    0    0    1
dp1$alpha :  6  3 -6 -3 
mix.p : 1 
\end{verbatim}
\begin{table}[ht]
\centering
\caption{Error quantiles for a fixed grid of points} 
\begin{tabular}{lllllll}
  \hline
 & 25\% & 50\% & 75\% & 90\% & 95\% & 99\% \\ 
  \hline
kde & 0.00198 & 0.0159 & 4.86E+11 & 3.75E+51 & 3.97E+82 &  Inf \\ 
  wkde & 0.00198 & 0.0159 & 4.72E+11 & 3.57E+51 & 3.75E+82 &  Inf \\ 
  wkdeA & 0.00198 & 0.016 & 4.82E+11 & 3.66E+51 & 3.84E+82 &  Inf \\ 
  fill+wkde & 0.00204 & 0.0188 & 3.69E+12 & 1.61E+56 & 4.42E+90 &  Inf \\ 
   \hline
\end{tabular}
\end{table}
\begin{table}[ht]
\centering
\caption{Error quantiles evaluating at the observed sample points} 
\begin{tabular}{lllllll}
  \hline
 & 25\% & 50\% & 75\% & 90\% & 95\% & 99\% \\ 
  \hline
kde & 0.0265 & 0.0576 & 0.11 & 0.198 & 0.292 & 0.701 \\ 
  wkde & 0.0265 & 0.0578 & 0.111 &  0.2 & 0.294 & 0.704 \\ 
  wkdeA & 0.0266 & 0.0579 & 0.111 & 0.199 & 0.293 & 0.701 \\ 
  fill+wkde & 0.0241 & 0.0524 & 0.101 & 0.184 & 0.273 & 0.652 \\ 
   \hline
\end{tabular}
\end{table}
\clearpage\centerline{\Large Case No.\, 54 }
\begin{verbatim}
Niter : 2500 
n : 500 
d : 4 
m : 3 
qN : 3 
Npts : 2401 
dp1$xi : 0 0 0 0 
dp1$Omega : 
         [,1]     [,2]     [,3]     [,4]
[1,]  1.00000  0.60714 -0.04464 -0.33705
[2,]  0.60714  1.00000  0.60714 -0.04464
[3,] -0.04464  0.60714  1.00000  0.60714
[4,] -0.33705 -0.04464  0.60714  1.00000
dp1$alpha :  6  3 -6 -3 
dp1$nu : 2 
mix.p : 1 
\end{verbatim}
\begin{table}[ht]
\centering
\caption{Error quantiles for a fixed grid of points} 
\begin{tabular}{lllllll}
  \hline
 & 25\% & 50\% & 75\% & 90\% & 95\% & 99\% \\ 
  \hline
kde & 0.000443 & 0.00242 & 0.0128 & 0.0505 & 0.105 & 0.374 \\ 
  wkde & 0.000443 & 0.00243 & 0.0128 & 0.0506 & 0.106 & 0.373 \\ 
  wkdeA & 0.000443 & 0.00243 & 0.0128 & 0.0506 & 0.106 & 0.373 \\ 
  fill+wkde & 0.000444 & 0.00243 & 0.0128 & 0.0507 & 0.106 & 0.374 \\ 
   \hline
\end{tabular}
\end{table}
\begin{table}[ht]
\centering
\caption{Error quantiles evaluating at the observed sample points} 
\begin{tabular}{lllllll}
  \hline
 & 25\% & 50\% & 75\% & 90\% & 95\% & 99\% \\ 
  \hline
kde & 0.0348 & 0.085 & 0.189 & 0.344 & 0.556 & 4.94 \\ 
  wkde & 0.035 & 0.0854 & 0.19 & 0.346 & 0.56 & 4.99 \\ 
  wkdeA & 0.035 & 0.0854 & 0.19 & 0.346 & 0.561 & 4.99 \\ 
  fill+wkde & 0.0348 & 0.0849 & 0.189 & 0.344 & 0.555 & 4.94 \\ 
   \hline
\end{tabular}
\end{table}
\clearpage
\small
\begin{verbatim}
Summary: relative improvement of error quantiles with Plain variant
rel.improvP : 
   d m  mix.p    n    grid.50     obs.50    grid.75     obs.75    grid.95     obs.95
1  4 3 0.3333  500 -7.024e-06  0.0062091  6.057e-05  0.0035047  0.0014075  0.0023555
2  4 2 0.3333  500 -4.816e-04  0.0632107  4.517e-04  0.0496820  0.0281589  0.0487456
3  4 3 0.3333  500  8.854e-04  0.0010320  4.857e-03  0.0007320  0.0064476  0.0001832
4  4 2 0.3333  500  3.086e-02  0.0075780  5.299e-02  0.0066434  0.0596355  0.0062920
5  4 3 0.3333  500  9.247e-05 -0.0019726 -5.787e-04 -0.0034900 -0.0056127 -0.0049934
6  4 2 0.3333  500 -1.073e-02  0.1587915 -8.131e-03  0.0766432  0.1658389  0.0535351
7  4 3 0.6667  500  9.290e-05 -0.0025718 -4.604e-04 -0.0024352 -0.0039162 -0.0043010
8  4 2 0.6667  500 -1.272e-02  0.1035375 -6.344e-03  0.0723508  0.1285367  0.0483129
9  4 3 1.0000  500 -5.496e-04 -0.0061320 -2.988e-03 -0.0078146 -0.0067139 -0.0068770
10 4 2 1.0000  500 -7.959e-03  0.0138169  1.543e-02  0.0400199  0.1232558  0.1119632
11 4 3 1.0000  500 -1.523e-02 -0.0115403  4.932e-03 -0.0129228  0.0199083  0.0013560
12 4 3 1.0000  500  4.819e-04  0.0001732  2.682e-02 -0.0115994  0.1283942 -0.0057434
13 4 2 1.0000  500  4.079e-02  0.0287409  3.354e-01  0.0208001  0.3082178  0.0502174
14 4 3 1.0000  500  1.392e-02 -0.0005857  1.794e-02 -0.0003149  0.0857899  0.0052793
15 4 2 1.0000  500  1.780e-01  0.0549584  1.853e-01  0.0490660  0.3811008  0.0629187
16 4 3 1.0000  500  1.549e-02  0.0183119  2.561e-02  0.0132881  0.0247188  0.0042106
17 4 2 1.0000  500  1.892e-01  0.1097304  3.741e-01  0.0922566  0.3969550  0.0870600
18 3 2 1.0000  500  5.556e-02  0.0653200  9.482e-02  0.0619619  0.3163063  0.0641892
19 3 2 1.0000  500  1.278e-01  0.0616721  1.092e-01  0.0616103  0.1263927  0.0748406
20 3 2 1.0000  500  7.405e-02  0.0116158  5.613e-02  0.0130387  0.1171125  0.0287566
21 3 2 1.0000  500  1.212e-01  0.0649599  1.197e-01  0.0632092  0.1219351  0.0703953
22 3 2 1.0000  500 -3.033e-02  0.1249595  2.755e-02  0.0842165  0.0500413  0.0997428
23 3 2 1.0000  500  1.408e-02  0.1458633  1.316e-01  0.1193372  0.1957737  0.0798755
24 3 2 1.0000  500  1.280e-02  0.1119781  5.674e-02  0.0825847  0.1378309  0.0490871
25 5 2 1.0000  500  2.696e-01  0.1640928  5.246e-01  0.1346124  0.3636144  0.0941541
26 5 3 1.0000  500  4.909e-02  0.0370564  1.004e-01  0.0243688  0.1197060  0.0004601
27 5 2 0.6667  500 -7.520e-02 -0.0304363 -2.542e-02 -0.0044766  0.1318902  0.1014228
28 5 3 0.6667  500  1.255e-03 -0.0060955  6.055e-04 -0.0053435  0.0031362 -0.0110032
29 5 2 0.6667  500 -6.223e-04  0.0587803 -5.015e-05  0.0519677  0.0112995  0.0727365
30 5 3 0.6667  500 -6.200e-05  0.0058907 -1.652e-05  0.0041479  0.0007639  0.0044376
31 5 2 0.6667  500  7.981e-02  0.0089196  8.742e-02  0.0057118  0.2295232  0.0057754
32 5 3 0.6667  500  8.174e-03  0.0005249  9.317e-03  0.0005979  0.0327445 -0.0004954
33 5 3 1.0000 1000  1.913e-02  0.0719979  1.118e-01  0.0487230  0.1075626  0.0061596
34 5 3 1.0000  250  2.444e-02  0.0129532  4.412e-02  0.0070907  0.0300844 -0.0081575
35 4 3 1.0000  250 -4.661e-03 -0.0003045  2.770e-03 -0.0085057  0.0155507 -0.0066655
36 4 2 1.0000  250  1.988e-03  0.0306898  2.359e-02  0.0336224  0.0848010  0.0593612
37 4 2 1.0000 1000  1.177e-03  0.0211555  5.332e-02  0.0386698  0.1650018  0.1494835
38 4 2 0.3333  250 -2.769e-04  0.0349276  6.568e-04  0.0278809  0.0361631  0.0305796
39 4 2 0.3333  250 -1.565e-03  0.1122842  6.472e-03  0.0603346  0.1604286  0.0743937
40 4 2 1.0000  250  2.481e-02  0.0342828  1.827e-01  0.0243983  0.2413600  0.0253359
41 4 2 1.0000 1000  3.731e-02  0.0142255  2.865e-01  0.0088516  0.2804137  0.0682228
42 4 2 1.0000  250  1.553e-01  0.0793144  2.639e-01  0.0659200  0.1952579  0.0646610
43 4 2 1.0000 1000  1.352e-01  0.1350389  4.624e-01  0.1144852  0.1098095  0.1037914
44 5 2 1.0000 1000  1.213e-01  0.2117341  5.973e-01  0.1692204  0.4616741  0.1138866
45 5 2 1.0000  250  1.976e-01  0.1174394  3.706e-01  0.0972674  0.2576898  0.0604635
46 4 2 0.3333  250  7.136e-03  0.0008641  2.205e-02  0.0008915  0.1833463 -0.0002455
47 4 2 0.3333 1000  5.672e-02  0.0173900  5.416e-02  0.0152593  0.0917034  0.0135010
48 4 2 1.0000  250 -1.282e-02  0.1762382  5.927e-02  0.1358727  0.2485566  0.0901304
49 4 2 1.0000  500 -2.638e-02  0.2399888  3.831e-02  0.1777767  0.3134105  0.0873002
50 4 2 1.0000 1000 -3.982e-02  0.2860935  3.176e-03  0.1967130  0.3255958  0.0511283
51 4 3 1.0000  250 -5.343e-04  0.0023868 -1.210e-04 -0.0022015 -0.0012113 -0.0157336
52 4 3 1.0000  500 -5.069e-04  0.0008678 -1.077e-03 -0.0087687 -0.0101262 -0.0329717
53 4 3 1.0000  250  7.377e-05 -0.0036595  2.863e-02 -0.0080380  0.0559315 -0.0069071
54 4 3 1.0000  500 -8.766e-04 -0.0047209 -1.407e-03 -0.0039813 -0.0012095 -0.0084011
\end{verbatim}
\clearpage
\begin{verbatim}
Summary: relative improvement of error quantiles with Fill variant
rel.improvF : 
   d m  mix.p    n    grid.50     obs.50    grid.75     obs.75    grid.95     obs.95
1  4 3 0.3333  500 -4.600e-04  0.0065579 -6.411e-03  2.335e-03 -6.859e-02  0.0022439
2  4 2 0.3333  500 -1.959e-02  0.0649588 -6.643e-02  3.934e-02 -1.079e+00  0.0440172
3  4 3 0.3333  500 -1.399e+01  0.0004061 -9.173e+00  6.291e-05 -6.550e+06 -0.0002471
4  4 2 0.3333  500 -1.160e+04  0.0027459 -3.888e+03  1.721e-03 -3.067e+14  0.0031734
5  4 3 0.3333  500  9.827e-04  0.0007644  1.951e-05 -2.568e-04 -2.826e-03  0.0008446
6  4 2 0.3333  500  6.021e-03  0.1741547  7.287e-04  8.196e-02  6.284e-02  0.1145357
7  4 3 0.6667  500  8.073e-04  0.0008974  1.189e-03 -2.243e-04 -2.716e-03  0.0003218
8  4 2 0.6667  500  1.550e-02  0.1657329  2.504e-03  8.309e-02  4.552e-02  0.1147171
9  4 3 1.0000  500  1.076e-03  0.0022720  1.565e-03  1.678e-03  2.173e-03  0.0033365
10 4 2 1.0000  500 -8.867e-01  0.3168617 -1.270e+00  3.677e-01 -6.757e-01  0.3515219
11 4 3 1.0000  500 -1.976e-01  0.1761395 -4.734e-01  1.718e-01 -2.449e-01  0.1259928
12 4 3 1.0000  500 -2.617e-01  0.1002917 -5.022e+01  9.682e-02 -1.329e+10  0.0910333
13 4 2 1.0000  500 -8.144e-01  0.1998363 -8.004e+03  2.178e-01 -1.122e+15  0.2994834
14 4 3 1.0000  500 -3.064e+00 -0.0013903 -1.199e+01  3.579e-03 -2.836e+06  0.0224118
15 4 2 1.0000  500 -2.003e+02  0.0315370 -2.827e+05  5.117e-02 -1.488e+15  0.1669600
16 4 3 1.0000  500 -1.805e+00  0.0184925 -1.562e+01  1.698e-02 -4.247e+05  0.0248494
17 4 2 1.0000  500 -3.474e+01  0.1123845 -2.280e+06  1.026e-01 -2.870e+17  0.1781653
18 3 2 1.0000  500 -2.543e+00  0.0673693 -2.094e+00  7.298e-02 -1.039e+05  0.1014109
19 3 2 1.0000  500 -2.745e+00  0.0444310 -1.163e+00  5.524e-02 -4.409e+00  0.1023358
20 3 2 1.0000  500 -1.717e+01 -0.0085571 -5.043e+02 -7.551e-04 -4.679e+08  0.0422174
21 3 2 1.0000  500 -6.003e+00  0.0447325 -1.435e+00  5.213e-02 -2.967e+00  0.0926267
22 3 2 1.0000  500  1.853e-02  0.2906052  5.808e-02  2.419e-01  1.503e-01  0.2865112
23 3 2 1.0000  500  2.427e-02  0.1115106  1.054e-02  8.790e-02  7.459e-02  0.1522683
24 3 2 1.0000  500 -3.636e-01  0.0895953 -3.814e-01  4.749e-02 -1.092e+00  0.0648791
25 5 2 1.0000  500 -7.552e+03  0.2255596 -4.661e+08  2.350e-01 -1.918e+18  0.3047575
26 5 3 1.0000  500 -5.164e+01  0.0770386 -5.662e+04  8.528e-02 -8.533e+10  0.0845638
27 5 2 0.6667  500 -3.261e-02  0.0011853 -2.148e-02  3.002e-03 -9.525e-03  0.1642382
28 5 3 0.6667  500 -1.229e-02 -0.0049056 -4.794e-03 -5.097e-03 -6.760e-03  0.0112358
29 5 2 0.6667  500 -4.957e-02  0.0695890 -1.461e-01  5.281e-02 -2.013e+00  0.0838223
30 5 3 0.6667  500 -7.283e-03  0.0087676 -2.375e-02  4.969e-03 -2.364e-01  0.0088703
31 5 2 0.6667  500 -1.716e+09  0.0031929 -2.377e+13  7.005e-04 -1.561e+23  0.0033200
32 5 3 0.6667  500 -5.526e+04 -0.0007819 -2.782e+06 -6.267e-04 -2.883e+13 -0.0011976
33 5 3 1.0000 1000 -9.046e+01  0.1110201 -7.476e+05  1.119e-01 -6.829e+13  0.1251793
34 5 3 1.0000  250 -1.906e+01  0.0447503 -6.994e+02  5.600e-02 -1.127e+09  0.0475500
35 4 3 1.0000  250 -1.403e-01  0.1631826 -1.835e-01  1.248e-01 -1.843e-01  0.0918364
36 4 2 1.0000  250 -5.069e-01  0.3671602 -6.925e-01  3.680e-01 -5.919e-01  0.3108264
37 4 2 1.0000 1000 -2.412e+00  0.2654077 -1.998e+00  3.256e-01 -1.236e+00  0.3468042
38 4 2 0.3333  250 -2.260e-02  0.0378355 -6.597e-02  2.369e-02 -9.244e-01  0.0312826
39 4 2 0.3333  250  3.576e-04  0.1308397 -2.349e-02  6.312e-02 -1.320e-01  0.1185749
40 4 2 1.0000  250 -7.377e-01  0.2406346 -1.997e+02  2.695e-01 -1.333e+11  0.2652797
41 4 2 1.0000 1000 -1.427e+00  0.1570527 -3.692e+03  1.433e-01 -2.168e+17  0.2897565
42 4 2 1.0000  250 -2.933e+01  0.0927426 -5.400e+03  9.286e-02 -2.071e+13  0.1600587
43 4 2 1.0000 1000 -3.373e+01  0.1223061 -2.937e+06  1.047e-01 -1.181e+20  0.1685462
44 5 2 1.0000 1000 -1.053e+04  0.2498730 -3.070e+10  2.472e-01 -2.087e+21  0.3463910
45 5 2 1.0000  250 -8.504e+03  0.1943648 -1.067e+07  2.217e-01 -1.141e+16  0.2580643
46 4 2 0.3333  250 -2.347e+02 -0.0037129 -3.010e+04 -3.090e-03 -1.462e+12 -0.0025253
47 4 2 0.3333 1000 -1.179e+04  0.0110207 -9.264e+03  8.527e-03 -3.248e+19  0.0089931
48 4 2 1.0000  250 -2.107e-01  0.2459490 -6.981e-01  2.060e-01 -6.854e-01  0.3602301
49 4 2 1.0000  500 -3.605e-02  0.3055664 -1.901e-01  2.454e-01 -2.352e-01  0.4245999
50 4 2 1.0000 1000  4.666e-02  0.3433847  5.629e-02  2.699e-01  2.940e-01  0.4479900
51 4 3 1.0000  250  1.637e-02  0.0127077 -1.676e-02  9.533e-03 -3.566e-02  0.0099689
52 4 3 1.0000  500  1.214e-02  0.0172478 -1.972e-03  1.108e-02 -2.081e-02  0.0128858
53 4 3 1.0000  250 -1.813e-01  0.0893326 -6.598e+00  8.483e-02 -1.112e+08  0.0658840
54 4 3 1.0000  500 -2.634e-03  0.0006553 -5.151e-03  3.109e-04 -1.835e-03  0.0011444
\end{verbatim}

\end{document}